# Entitled to Property: How Breaking the Gender Barrier Improves Child Health in India[†]

Md Shahadath Hossain[a]    Plamen Nikolov [bcd]

**Abstract.** Non-unitary household models suggest that enhancing women's bargaining power can influence child health, a crucial determinant of human capital and economic standing throughout adulthood. We examine the effects of a policy shift, the Hindu Succession Act Amendment (HSAA), which granted inheritance rights to unmarried women in India, on child health. Our findings indicate that the HSAA improved children's height and weight. Furthermore, we uncover evidence supporting a mechanism whereby the policy bolstered women's intra-household bargaining power, resulting in downstream benefits through enhanced parental care for children and improved child health. These results emphasize that children fare better when mothers control a larger share of family resources. Policies empowering women can yield additional positive externalities for children's human capital. (*JEL* D13, I12, I13, J13, J16, J18, K13, O12, O15, Z12, Z13)

*Keywords*: human capital, height, bargaining, parental investments, developing countries, India

[†]We thank Charlotte Williams, Andrew Ryan, Declan Levine, Matthew Bonci, Justin Sun, and Ariunzaya Oktyabri for exceptional research assistance. We thank Kirill Borusyak, Denni Tommasi, Julie Cullen, Susan Godlonton, Stephen Gire, Eric Edmonds, Lauren Bergquist, Rachel Heath, Agnitra Roy Choudhury, Paul Novosad, Angela de Oliveira, Lisa Kahn, Seema Jaychandran, Belinda Archibong, Sulagna Mookerjee, Livia Montana, Jessica Goldberg, S. Anukriti, Petra Moser, and Evan Riehl for constructive feedback and helpful comments.

*Corresponding Authors: Hossain: Department of Economics, University of Houston, Department of Economics, Science Building, 3581 Cullen Boulevard Suite 230, Houston, TX 77204-5019, USA (e-mail: mhossai3@binghamton.edu); Nikolov: Harvard IQSS, 1737 Cambridge Street, Cambridge, MA 02138, USA (e-mail: pnikolov@post.harvard.edu)

[a] University of Houston
[b] IZA Institute of Labor Economics
[c] Harvard Institute for Quantitative Social Science
[d] Global Labor Organization

# I. Introduction

Height in early childhood strongly predicts cognitive ability, educational attainment, occupational choice, and labor market outcomes (Case and Paxon, 2010; Persico, Postlewaite, and Silverman, 2004). Stunting, or low height-for-age, affects about 25 percent of children under five in low- and middle-income countries (LMICs). A quarter of these global stunting cases are reported from India, where the prevalence rate stands at 31 percent, according to UNICEF's 2020 data.

Several factors, such as genetics, biology, and disease environment, contribute to a child's height in early life.[1] However, resource allocation at the household level also plays a significant role (Rosenzweig and Schultz, 1982; Jayachandran and Pande, 2017). The distribution of these resources often depends on control within the family, particularly women's bargaining power (Thomas, 1990). Enhancing female bargaining power can improve family well-being and mitigate gender inequality. Further, the 'maternal altruism' hypothesis argues that women, when they have control over resources, are more likely than men to invest in children and less likely to favor boys. Maitra (2004) shows that women allocate more resources to nutrition, medical care, and childcare, a finding in line with the 'maternal altruism' hypothesis that women are more inclined to invest in children than men and less likely to favor boys (Dizon-Ross and Jaychandran, 2022).[2] Historically, male-biased inheritance practices have constrained women's land ownership and bargaining power in India, negatively influencing household spending patterns and potentially compromising children's outcomes.

In this study, we explore the effect of the passage of inheritance law amendments, which conferred improved inheritance rights to unmarried women in India[3], on child height and weight. In India, a predominantly rural country, land ownership is a critical determinant of economic and social status (Chowdhry, 2017); inheritance is the primary way to acquire such land. However, conventional male-biased inheritance practices and cultural conservatism (Chowdhry, 2017) engender limited earning opportunities and bargaining power for women. Because of the potential to transform women's inheritance rights, some Indian states amended the male-favored Hindu Succession Act (from now on, *HSA*) to provide unmarried women an equal share in ancestral property. Improved female bargaining power could enhance human capital outcomes and gender inequality. Thus, enhanced inheritance rights for women may lead to significant downstream benefits: better human capital outcomes and improved status of daughters. We explore this possibility in this paper.

---

[1] Several studies examine the role of genetic factors (Rootsi et al., 2004) and disease environment (Bozzoli, Deaton, Quintana-Domeque, 2009). Other studies focus on cultural gender preference and unequal resource allocations within a family based on perceived returns on the investment (Rosenzweig and Schultz 1982; Oster 2009; Jayachandran and Pande 2017).

[2] Given these different spending patterns, attention to better female bargaining power at the household level can improve girls' survival rates and anthropometric measures (Duflo 2012).

[3] The amendment only affected unmarried women at the time of the enactment.

Our estimation approach relies on the staggered adoption of the HSA amendments (from now on, *HSAA*) across different states in India since 1956. We rely on two primary sources of variation related to a woman's exposure to the HSAA. The first source is the timing of a woman's marriage. The reform in each state only affects unmarried women; thus, the policy does not affect women who marry before the HSAA.[4] The second source of variation stems from the HSAA implementation: some states amended the HSA before the 2005 federal amendment. The treated states in our sample are Andhra Pradesh, Tamil Nadu, Karnataka, and Maharashtra. We exploit these quasi-experimental variations and use a new difference-in-differences (*DiD*) estimator developed by Borusyak et al. (2021) for settings with staggered adoption of treatment and heterogeneous treatment effects. An alternative to the conventional event study model, this new estimator offers more desirable robustness and improved efficiency properties (Roth et al., 2023). We implement the method using data from the Indian National Family and Health Survey (NFHS), a cross-sectional survey with three rounds of data: Wave 1 (1992-1993), Wave 2 (1998-1999), and Wave 3 (2005-2006).

We find clear evidence that the HSAA substantially improved child health. The HSAA improved child height, or height-for-age (HFA), by 0.147 standard deviations (SDs). Similarly, we detect a large effect on weight, proxied by the weight-for-age (WFA) measure, by 0.272 SDs. As a robustness check, we present conventional two-way fixed effect estimates showing similar effects on child height and weight. To examine the source of these improvements, we analyze the impact of the HSAA on parental investments. In terms of heterogeneous effects, we find evidence that the HSAA positively affects the daughters' health, but only on short-term health outcomes, such as weight. We find no evidence that the HSAA improved their height or parental care. Turning to higher birth-order children, the HSAA has no positive impact and may even have an adverse effect by reducing parental allocations, thereby impacting their health negatively. Lastly, we find no evidence of any favorable impact of the HSA reform on daughters of higher birth order.

---

[4] One caveat could be that parents can purposefully advance (those who want to avoid devolving property to daughters) or delay (those who are gender progressive) daughters' marriage. This sort of selection based on the marriage timing is not necessarily a concern as the amendments were often implemented retrospectively. We account for the age at marriage to account for any potential bias.

Our findings provide strong evidence of the HSAA's positive impact on child health, but these benefits are limited to first-born and second-born children. In India, where a preference for first-born sons influences household resource allocation, larger families tend to allocate fewer resources to later-born children, especially daughters.[5] Previous research highlights that factors such as resource constraints and sibling rivalry, compounded by the fertility-stopping effect, place later-born daughters at a disadvantage (Jayachandran and Pande, 2017).[6,7] Therefore, we specifically examine the HSAA's effect on the health of daughters and later-born children, but we find no evidence of improved health outcomes for later-born daughters.

Furthermore, we investigate the underlying mechanisms that contribute to the positive health effects of the HSAA. Our empirical analysis reveals a connection between the HSAA and increased bargaining power and female empowerment within households in the implementation areas.[8] Women in these areas report higher land ownership, marriage to better-educated and wealthier husbands, and reduced dowry payments. Additionally, we examine the impact of the HSAA on indicators of female empowerment, such as women's influence in significant household purchases, healthcare decision-making, and the ability to travel independently to the market. Our findings robustly demonstrate that the HSAA significantly enhances female bargaining power.[9]

---

[5] The patrilineal system promotes a strong preference for sons because they typically live with their parents, take care of them in their old age, work on the family land, and eventually inherit it. On the other hand, daughters marry outside their birth home and receive family assets as dowry (Rosenzweig and Schulz 1982). There is existing evidence that son-preference is stronger in families where the first-born child is not a son; these families tend to be larger as they keep growing until their desired gender mix of children is achieved (Arnold, Choe, and Roy, 1998).

[6] "Fertility stopping" refers to the phenomenon of families with daughters only, in which the parents conserve resources in anticipation of a male child.

[7] A daughter with no elder brother may benefit from a lack of sibling rivalry. Still, the fertility-stopping effect reduces resources received by the daughter as the family realizes that they need to try again for a son and start saving funds for their next child (an expected son). Jayachandran and Pande (2017) find that the fertility-stopping effect dominates, and a daughter with no brother receives relatively fewer postnatal health inputs than prenatal inputs.

[8] Building on Heath and Tan's (2020) findings on the HSAA's effect on women's labor and income, we highlight its key role in enhancing their decision-making and bargaining clout.

[9] The finding is consistent with Calvi (2020), a study showing that the HSAA increased female bargaining power in household decisions. Moreover, Lundberg, Pollak, and Wales (1997) test the non-unitary household model and conclude that improved female bargaining power can enhance children's outcomes. In line with this discovery, Duflo (2012) reviews the empirical evidence specifically with data from LMICS and highlights the differential impact of income in the hands of women versus men on intra-household allocation.

Our study contributes to the empirical literature on female empowerment's impact on household economic outcomes in LMICs. First, our findings substantiate the non-cooperative bargaining models of marriage (Lundberg and Pollak, 1993), which posit that within marital relationships, individuals act in their own self-interest and engage in strategic decision-making to maximize their personal benefits and outcomes. Land ownership and changes to inheritance laws can significantly impact bargaining power within non-cooperative bargaining models of marriage, as they alter the distribution of economic resources and the ability of individuals, particularly women, to negotiate favorable outcomes and assert their preferences within the marital relationship. In developing countries, where land ownership plays a significant role, the relevance of changes in inheritance laws cannot be understated. Thus, we contribute to the existing literature on non-cooperative bargaining models of marriage by establishing a crucial link between enhanced female empowerment, facilitated by improved inheritance rights, and tangible health outcomes. Second, we uncover cautionary evidence that increased female bargaining power does not guarantee better human capital outcomes for all children within a household. These results challenge some "maternal altruism" models that suggest increased women's bargaining power would lead to better outcomes for all children (Duflo, 2012). Birth order emerges as a potentially significant factor, with no evident positive impact of the HSAA on later-born children's health outcomes. Our final contribution highlights the profound downstream influence of enhanced female empowerment on a critical child welfare measure: height. Despite the myriad factors contributing to stunting, we present compelling evidence that enhanced female empowerment can significantly bolster child height in LMICs. This insight aligns with past research emphasizing better child outcomes when women wield increased control in the household (Lundberg et al., 1997).

The rest of the paper is structured as follows. Section II provides background on the Hindu Succession Act Amendment and evidence of the reform's impact on female bargaining. Section III describes the data. Section IV outlines the empirical strategy. Section V reports and discusses the results. Section VI concludes.

## II.     Background

## A. *The Hindu Succession Act 1956, the 2005 Amendment, and the Impact on Women's Bargaining*

In the 1950s, India ushered in a wave of legal changes, culminating in the Hindu Succession Act (HSA) of 1956. This Act's primary purpose was to consolidate the two distinct inheritance systems in place and elucidate women's inheritance rights to private property

Before the HSA in 1956, two different systems guided property inheritance in India: *Dayabhaga* (in West Bengal and Assam) and *Mitakshara* (the rest of India) (Chowdhry, 2017). The former, authored by the Indian Sanskrit scholar Jimutavahana, established inheritance laws for the Hindu community and was the principal authority in the Bengal region's British Indian courts. Meanwhile, the rest of India adhered to the Mitakshara. Known for its measured approach, the Mitakshara held sway in Bengal on all legal matters where no conflicting Dayabhaga perspectives arose.

The main distinction between these two systems was categorizing and distinguishing between private (separate) and joint family (or ancestral) property.[10] Personal property is generally self-acquired and cannot pass via patrilineal succession (i.e., from a male line, such as a father or a grandfather). On the other hand, ancestral or joint family property is commonly inherited patrilineally. The *Dayabhaga* system did not distinguish between the two types of properties. Under the *Dayabhaga* system, all heirs, including sons, daughters, or widows, could claim to inherit property. By comparison, the *Mitakshara* system distinguished the inheritance procession as dependent upon whether the property was private or joint-family. Under the system, the owner can bequeath privately owned property to anyone; however, only *co-parceners* (i.e., sons, grandsons, or great-grandsons) could inherit joint family property.

The HSA aimed to promote gender equity by conferring women equal rights over private property. However, the Act did not apply to joint family property. Therefore, under the HSA, daughters had an equal right to their father's private property if the Hindu (i.e., Hindus, Buddhists, Jains, and Sikhs) male died without a will *(intestate)*.[11] However, women could not inherit joint family property. In contrast, sons being co-parceners had the right to inherit the joint family property by birth, which implied that their share of the property could not be willed away. They could even demand a division of the joint family property if older co-parceners were alive. Therefore, because of its different treatment of men and women who could inherit joint family property, the HSA discriminated against women.

---

[10] "*Joint family*," a legal term, does not require that people share the same household. "Joint family property" consists of property principally inherited via patrilineal succession (e.g., father, paternal grandfather, or paternal great-grandfather), plus property jointly acquired by the joint family or property acquired separately but merged into the joint family property. In India, most household wealth comprises joint family property (Chowdhry, 2017).

[11] Chowdhry (2017) indicate that intestacy occurs in roughly 65% of Indian deaths, with a higher incidence in rural regions.

The federal and state governments had legislative authority over inheritance issues in India (Roy, 2015).[12] Although HSA was a federally mandated law, states could pass amendments that guided the rules within the states' jurisdictions. For instance, Kerala[13] (in 1976), Andhra Pradesh (in 1986), Tamil Nadu (in 1989), Karnataka (in 1994), and Maharashtra (in 1994) passed state amendments to the HSA. The amendments granted Hindu women equal inheritance rights to joint family property, provided they were unmarried at the time of the amendment.[14]

The HSAA, in the *Mitakshara* system, elevated the status of daughters to that of sons, but these benefits pertained to unmarried daughters. Daughters who married before the commencement of the HSAA could not claim coparcenary status in the HSAA. This aspect of the law implies that married daughters do not enjoy the same rights on coparcenary property as sons. There are three reasons why the law excludes married daughters. First is dowry practice. Second, in some communities (such as *Kammas*), daughters receive a share of the property when they marry. However, the dowry practice is technically illegal in India. Only a few communities practice giving property at the time of marriage. Thus, a blanket exclusion of married daughters based on unlawful acts and practices in a few exceptional communities cannot be a reasonable justification (Sivaramayya, 1988).

In 2005, India enacted a nationwide extension of the Hindu Succession (Amendment) Act (HSAA) to address gender inequality in inheritance rights. The amendment modified Section 6, initially allowing gender discrimination in claiming joint family property inheritance. Following this amendment, the Supreme Court decreed the law's retroactive applicability. Consequently, a daughter gained co-sharing rights with her male siblings, regardless of whether the father was alive as of September 9, 2005 - the date of the amendment's enactment.

---

[12] The HSA applied nationwide, excluding Jammu and Kashmir which enacted its unique version. Despite special provisions for matrilineal societies, it excluded northeastern tribal communities that, although matrilineal, adhered to local customs (Chowdhry, 2017).
[13] The reforms in Kerala were quite distinct from the other state-level reforms as they abolished the system of joint property altogether (Roy, 2015).
[14] The amendments were often implemented retrospectively (Roy, 2015). For example, Andhra Pradesh formally passed the amendment in May 1986, but it was retrospectively in effect from September 1985 onwards.

### III. Data and Summary Statistics

#### A. *Child Health in India*

Despite India's remarkable economic progress in recent decades, this growth has not effectively translated into improved health outcomes, leaving India significantly lagging behind other developing countries in its health indicators (Jayachandran and Pande, 2017). Empirical research confirms that child health profoundly influences adult economic and health outcomes. Consequently, many recent studies delve into the reasons behind India's disconcertingly low average child height.

In rural India, the average child under five exhibits a height deficit of approximately two standard deviations compared to the World Health Organization's healthy growth reference population. Thirty-one percent of Indian children under five continue to experience stunting (UNICEF, 2020). It is worth noting that India, with a GDP per capita surpassing 60 countries, possesses the fifth-highest stunting rate globally. Even after accounting for maternal health and parental socioeconomic disparities, comparative analysis demonstrates that an average Indian child is more susceptible to stunting than a counterpart in Sub-Saharan Africa (Jayachandran and Pande, 2017).

There is also growing empirical evidence of the crucial role parental investments play in determining child height in India (Jayachandran and Pande, 2017). Differential parental care based on gender and birth order is a critical determinant of height. For instance, Fledderjohann et al. (2014) show that although infant girls and boys obtain the same caloric intake, families feed girls more cereals while giving boys more milk and fats.

However, enhancing female empowerment at the household level and implementing policy initiatives to improve female bargaining power can boost child health. When women are part of household decision-making, the resources they bring can impact the household's final decision-making. Extensive empirical research examines whether income in the hands of women has a different impact on intra-household allocation than income in the hands of men (Duflo, 2012). The evidence suggests that when more income is in the hands of women than men, there are significant subsequent improvements in child health and larger expenditure shares on household nutrients (Thomas, 1990). The theoretical model underlying these predictions is of a non-unitary household, a household as a collective of individuals with different preferences (Lundberg, Pollak, and Wales, 1997).

In predominantly rural nations like India, land, and family property inheritance rights can substantially augment women's empowerment. Economic and social status is significantly influenced by land ownership, with inheritance serving as the primary mode of acquisition (Chowdhry, 2017). Recognizing the transformative potential of such rights for women, select Indian states have amended the previously male-biased Hindu Succession Act (HSA), providing unmarried women equal entitlement to ancestral property. Consequently, improved female bargaining power could improve human capital outcomes and gender inequality.

In developing countries like India, high levels of human capital can serve as a pathway to escape poverty (Chakravarty et al., 2019). Consequently, fostering female empowerment holds the potential to generate beneficial spillover effects for the younger generation. Careful design and precise targeting are crucial to ensure the effectiveness of public policies aimed at promoting human capital accumulation. One specific policy measure that has shown promise is reinforcing inheritance rights for women, as it can potentially enhance human capital outcomes for children. In the subsequent sections of this paper, we delve into this premise using data from India.

### A. *The Data: A Descriptive Analysis*

*Indian National Family and Health Survey (NFHS).* For our analysis, we use data from the Indian National Family and Health Survey (NFHS), a cross-sectional survey comprising three rounds of data: Wave 1 (1992-1993), Wave 2 (1998-1999), and Wave 3 (2005-2006). The household survey covers rural and urban households. It adopts a stratified multistage cluster sampling method to identify a nationally representative sample of the population living in urban and rural areas in 29 states. The survey selected 110,000 households in each wave and collected information from 125,000 women (aged 15 to 49 years) and 75,000 men (aged 15 to 54 years).

All women of age 15 to 49 were eligible for an interview. However, because many health indicators pertain to the sample of ever-married women and children, the required sample size for men was considerably smaller. Thus, of the 216,969 eligible women and men, 124,385 women and 74,369 men participated in the survey, yielding a response rate of 94.5 percent and 87.1 percent, respectively.

The survey questionnaire comprises several distinct modules, including a household module, a module collecting information from women, and a village information module. The household module gathers information from all residents in each sample household via face-to-face interviews. The survey also covers demographic data on age, gender, marital status, relationship to the head of the household, education, and occupation for each listed person. Based on the household module, the survey team identified respondents eligible for the woman's questionnaire, collecting additional demographic information on adult female respondents and, when applicable, their children.

The woman's questionnaire collects information from all ever-married women aged 15–49 who were the residents of the sampled household. The module gathers background information on socioeconomic characteristics (age, marital status, education, employment status, place of residence), reproductive behavior (fertility choice, birth spacing, number of children, prenatal and postnatal healthcare use), and quality of childcare. The module also covers questions on all children (age, sex, birth order, and health information, such as height, weight, hemoglobin levels, and prior vaccinations). In addition, the survey gathers anthropometric measures for both adults and children. Finally, the NFHS collected height and weight measurements for children in all rounds.

Our analysis sample consists of 65,371 children for whom we have anthropometric data. We use data on the following fifteen states: the four states that implemented the HSAA (Andhra Pradesh, Tamil Nadu, Maharashtra, and Karnataka) and eleven states that did not implement the HSAA (Arunachal Pradesh, Bihar, Goa, Gujrat, Haryana, Himachal Pradesh, Madhya Pradesh, Orissa, Panjab, Rajasthan, and Uttar Pradesh).[15]

---

[15] Our analysis omits Union territories, West Bengal, Jammu and Kashmir, Kerala, and Northeastern states due to their unique political, administrative, and inheritance systems, and matrilineal kinship areas (Jayachandran and Pande 2017). Similar exclusions appear in other HSA-related studies (e.g., Rosenblum, 2015). Including these states, however, does not alter our conclusions (see Appendix A Table 11).

*Rural Economic and Development Survey (REDS).* In addition to NFHS, we use the 1999 wave of the Rural Economic and Development Survey (REDS) as the survey collects additional data on inheritance-related assets. REDS, a nationally representative household survey of 7,500 rural households in sixteen major states of India, collects data from the household head as the primary survey respondent. The head provides retrospective information on all household members, including daughters. The survey collects data on the daughter's age at marriage, cash dowry payment, daughter's land inheritance(s), and current land ownership. We use these variables for descriptive analyses to link the HSAA implementation with higher subsequent unearned income and better empowerment outcomes for women in these areas. Our primary focus is the married daughters of household heads above 15. In addition, we keep Hindu (Hindu, Buddhist, Sikh, and Jain) households subject to the HSAA. Finally, to be consistent with the NFHS criterion, we drop observations from Assam, West Bengal, and Kerala. The resulting sample data comprises thirteen states: the four states that implemented the HSAA (Andhra Pradesh, Tamil Nadu, Maharashtra, and Karnataka) and the nine states that did not implement the HSAA (Bihar, Gujrat, Haryana, Himachal Pradesh, Madhya Pradesh, Orissa, Panjab, Rajasthan, and Uttar Pradesh). Our analysis sample consists of 3,463 women for whom we have data on cash dowry payment, land inheritance, and land ownership. The average cash dowry payment is USD 37 (SD 0.089). Only 3.44 percent (SD 18.2 percent) of the women received a land inheritance and owned 0.195 (SD 0.561) acres.

### B. Study Outcomes

**Child Health**. We use height as an indicator of early childhood health. Height is a stock variable and a meaningful indicator of accumulated decisions regarding nutritional intake in early life (Case and Paxson, 2010). We also use weight as an additional proxy for child health. However, weight is a flow measure and captures short-term changes to the nutritional environment.

Based on the raw measurements for these variables, we create standardized measures for height and weight using additional information on age, height, and weight for all children. The two standardized measures for height and weight are height-for-age (HFA) and weight-for-age (WFA).[16] The HFA z-score is available for children under age five. The WHO defines children with two and three standard deviations lower than the mean HFA z-score as moderately and severely stunted (WHO, 2006). Similarly, children two and three standard deviations lower than the mean WFA z-score are defined as moderately and severely wasted, respectively.

We also examine other indicators of child health. We analyze several indicators, including the total number of prenatal visits, whether the mother received iron supplementation, whether she received a tetanus shot, whether the delivery was performed in a health facility, whether the family conducted a postnatal check within two months of birth, and the vaccination status of the child. Postnatal checks are only available for the youngest living child. We also create a composite normalized input index based on prenatal and postnatal inputs. This index helps us to gauge the quality of parental care for children. Furthermore, we measure disease incidence in the last two weeks to capture the early childhood disease environment. We do so by constructing a composite index based on the following variables: the incidence of the child having a fever, cough, or diarrhea in the last two weeks. Based on these variables, we create a normalized composite score.

**Female Bargaining Power Measures**. Besides the socioeconomic and health outcomes in the survey, we also explore outcomes related to childbearing and women's autonomy. The NFHS collects data on the number of children *ever born* and the mother's age at first marriage. We measure women's bargaining using three binary variables: whether the woman contributes to the decision-making of large purchases, whether the woman contributes to the decision-making guiding her own healthcare needs, and whether she can go to market alone. This approach tracks the one adopted by Heath and Tan (2020). Based on these indicators, we generate a composite normalized index for female bargaining power.

*C. Sample Summary Statistics*

Table 1 reports the summary statistics for the main variables. The table reports data on the entire sample, the sample in only states affected by the policy change (columns 2 through 4), and the sample only in states unaffected by the policy (columns 5 through 7). Primarily, the policy change affects Hindus, who define their family ties through a solid patrilineal and patrilocal kinship system. Therefore, we split the sample by religion: column 2 comprises the Hindus affected by the policy; column 3 reports data on non-Hindu individuals in the treated states. Columns 5-7 present the summary statistics for the sample (Hindus and non-Hindus) and non-treated states. Table 1 shows data based on the mothers' sample: age, age at first marriage, age at first birth, and various proxies of bargaining power.

---
[16] A z-score of 0 represents the median of the reference population, while a z-score of -2 indicates that the child is two standard deviations below the mean of the reference population.

[Table 1 about here]

In both treated and non-treated states, the average mother's age is around 17 at the time of her first marriage. Hindu mothers living in areas affected by the policy have similar ages at first marriage to mothers living in areas unaffected by the policy. Based on the index we construct, Hindu women in states affected by the HSAA exhibit higher bargaining power than those in states that did not adopt the HSAA.

Turning to the children's characteristics in the whole sample, Table 1 reports that the share of daughters as a proportion of all children is nearly identical between states who adopted the HSAA. The average HFA is -1.935, but higher in states that adopted the HSAA (i.e., -1.666); in states that did not, the HFA is -2.006. The summary statistics for the WFA measure present the same pattern.

## IV. Empirical Strategy

To estimate the effect of the HSA amendment on child health outcomes, we take advantage of the staggered adoption of the HSAA across states. We use two sources of identifying variation. The first source is the timing of a woman's marriage. The second source relates to whether the woman's state of residence adopted the 1956 HSAA; the reform only affected unmarried women when the reform occurred in their state.[17] Women who were married before the reform comprise the control group. Four states adopted the reform:

---

[17] A potential concern could be out-migration to different states from where the woman was born. However, since cultural and linguistic barriers impede cross-state migration in India, cross-state migration is negligible (Roy, 2015).

Andhra Pradesh, Tamil Nadu, Karnataka, and Maharashtra.[18] Eleven states did not: Arunachal Pradesh, Bihar, Goa, Gujrat, Haryana, Himachal Pradesh, Madhya Pradesh, Orissa, Punjab, Rajasthan, and Uttar Pradesh. All women from these non-reform states are also part of the control group, irrespective of their marital status.

Difference-in-differences (DiD) design with staggered adoption of treatment assumes that the true causal model for the outcome of interest reported in state $s$ in year $t$ is:

$$H_{ist} = \alpha_s + \beta_t + \tau_{st} D_{st} + \varepsilon_{ist} \qquad (1)$$

Here $H_{ist}$ is the outcome for child $i$ born to a woman who married in year $t$, in state $s$. $\alpha_s$ and $\beta_t$ capture the state and year (women's birth year) fixed effects.[19] $D_{st}$, the treatment indicator equals one if the state adopted the HSAA and the woman was unmarried before the enactment of the HSAA in that state. Furthermore, $\tau_{st}$ captures the treatment effect– that is, the impact of the HSAA on the child's health. And $\varepsilon_{ist}$ is the residual such that $E[\varepsilon_{ist}|\alpha_s, \beta_t, D_{st}] = 0$. The expected outcome of the model in (1) is $\alpha_s + \beta_t$ in the absence of treatment, where $\alpha_s + \beta_t$ captures the parallel trend assumption, and the model in (1) allows for heterogeneous treatment effects by state and year.

According to recent econometrics research (Roth et al., 2023), when there is effect heterogeneity, using a traditional two-way fixed effects event study design to estimate (1) may lead to unreliable and biased estimates. Several studies propose alternative estimators to address this issue. Although most of these estimation procedures are well-suited for panel datasets, the approach Borusyak et al. (2021) developed extends to both panel and repeated cross-sectional data settings. Using the repeated cross-sectional data, we rely on the estimator proposed by Borusyak et al. (2021).

The difference-in-differences (DiD) imputation approach relies on a framework that explicitly notes and isolates the underlying assumptions. This approach takes the standard DiD assumptions (i.e., parallel trend and no anticipation effects) and derives the optimal estimator for valid inference. The estimation approach is transparent, flexible (i.e., it allows for unit-specific trends, time-varying covariates, observational weights, and repeated cross-sections), and computationally efficient (i.e., it provides conservative standard errors).

---

[18] We do not use the national adoption of HSAA in 2005. Although it might seem that using national adoption of the HSAA will add variation to the current set of control states, it will lead to under-identification of the difference-in-differences (DiD) estimates. The under-identification problem arises from failing to strongly impose the no anticipation assumption of DiD estimation when there is no never-treated group (Borusyak et al., 2021). Once we move to all states being treated, the "no anticipation" assumption cannot be strictly enforced, which leads to under-identification.

[19] We use mothers' year of birth fixed effects to control for maternal age, as several studies have demonstrated a relationship between maternal age and the risks of child mortality and undernutrition (Finlay, Ozaltin, and Canning, 2011).

The DiD imputation estimator, following Borusyak et al. (2021), is constructed in three steps. First, the state and year fixed effects $\alpha_s$ and $\beta_t$ in equation (1) are estimated by OLS on the subsample of untreated observations only (i.e., $D_{st} = 0$). We estimate the state and year fixed effects using data on all women in non-reform states (i.e., states that did not adopt the HSAA) and women who were already married at the time of HSAA adoption in the reform states. Second, these state and year fixed effects are used to impute the untreated potential outcomes and therefore obtain an estimated treatment effect, $\hat{\tau}_{ist} = Y_{ist} - \hat{\alpha}_s - \widehat{\beta_t}$ for each treated observation. Finally, a weighted average of these treatment effect estimates is taken with weights corresponding to the estimation target. Our model estimates an average effect using a treatment horizon ($h$), denoting the years since treatment based on the following:

$$\hat{\tau}_h = \frac{1}{|I_h|}\sum_{i \epsilon I_h} \hat{\tau}_{s,E_s+h} \qquad (2)$$

$E_s$ is the year when state $s$ adopted HSAA, treatment horizon greater than or equal to zero ($h \geq 0$), and $I_h$ is the set of states $s$ observed in the period $E_s + h$ (i.e., in a year $h$ years after the HSAA). For each horizon, the imputation approach uses all difference-in-differences comparisons between some state $s$ in period $E_s + h$ relative to periods before HSAA and relative to other states which have not adopted HSAA yet by $E_s + h$.

The DiD imputation approach can address the violation of non-anticipatory effects by redefining the treatment indicator switch to one $K$ period before the treatment. Therefore, the validity of this estimation approach depends on the parallel trends assumption. Roth (2022) and Roth et al. (2023) summarize empirical evidence that due to varying treatment effects, pre-trend tests may reject the parallel trend assumption even when it is satisfied or pass the tests when the assumptions do not hold. To address this issue, Borusyak et al. (2021) take a different approach from the conventional method, which jointly estimates pre-trend or placebo coefficients with treatment effects. Instead, they conduct pre-trend tests using only untreated observations:

$$Y_{ist} = \alpha_s + \beta_t + \sum_{p=-P}^{-1} \gamma_p \mathbf{1}[t = E_s + p] + \varepsilon_{ist} \qquad (3)$$

Here, $\mathbf{1}[t = E_s + p]$ are indicator variables of being treated 1 to $P$ years later. We set $P = 8$, which implies having leads leading up to eight years. In this estimation, the control group comprises observations for which treatment happens more than $P$ years later. We can visually examine the magnitude of $\hat{\gamma}_p$ and conduct conventional individual and joint significance tests of $\gamma_p = 0$.

Borusyak et al. (2021) highlight three advantages of the approach. First, it separates the validation of the design (i.e., ex-ante parallel trend assumption) from the estimation step. This separation prevents the conflation of using an identifying assumption and validating it simultaneously. Second, it improves the efficiency of the treatment effect by imposing no pre-trend at the estimation stage. This efficiency gain arises from using all the untreated observations in the imputation. Finally, this approach resolves the issues Roth (2022) raises, as this estimation removes the correlation between the treatment effect and the pre-trend estimators.

We include the traditional two-way fixed effect estimation in Section V.F. as a reference point, in addition to the DiD imputation approach. We use this comparison to perform various robustness checks. Before formally estimating the above equations, we examine the trend in the outcome variables for the treated and non-treated areas. Figure 1 displays the trajectory of the HFA and WFA outcomes. The figure reveals an upward trend for both outcomes, regardless of treatment status. However, HFA in the treated states was higher than in the non-treated states between 1978 and 2004. On the other hand, the WFA is similar between the treated and non-treated states until 1989; after 1989, the gap between these two areas widens.

[Figure 1 about here]

## V. Results

### A. *Effects of the Reform on Child Health*

Based on equation (1), we present the HSAA's impacts on health and parental investment outcomes in Table 2. The table reports the $\hat{\tau}$ –the average treatment effect on the treated (ATT)–and standard errors in parentheses. We cluster the standard errors at the state level to address within-state correlations and heteroskedasticity between states.[20] All columns of Table 2 include control variable and fixed effects (FE)– state fixed effect and year fixed effect (also capturing variation in the maternal year of birth).

The results in Table 2 suggest that women-favored inheritance rights reform in India has significantly improved child height and weight. The estimates for health outcomes– height-for-age (HFA) z-score and weight-for-age (WFA) z-score– are presented in columns 1 and 2. The ATT in columns 1 and 2 are positive and statistically significant. For example, in column 1, the ATT is 0.147, indicating that the HSAA improves children's HFA z-score by 0.147 SDs. Similarly, the ATT in column 2 shows that the HSAA improves the WFA z-score by 0.272 SDs.

Table 2 also reports the HSAA effects on parental care for child health. The outcome variables, reported in columns 3-6, are prenatal care inputs. The outcome variables in columns 7-8 are the postnatal care inputs. We use four indicators of prenatal inputs during pregnancy: total prenatal visits, whether the mother took iron supplementation, whether the mother received a tetanus shot, and whether the delivery occurred at a health facility. Delivery at a health facility indicates whether the childbirth happened at a health facility instead of at home. Total prenatal visits and delivery at a health facility are available for all children under five. Data on iron supplement intake and tetanus shots are available only for the most recent birth (i.e., the youngest child). We measure postnatal parental inputs using two indicators: postnatal check within two months of birth and the total number of child vaccinations. Postnatal checks within two months of birth are available only for the most recent delivery (youngest child). Finally, the outcome variable, reported in the last column, pooled all inputs into an aggregated and standardized index of prenatal and postnatal inputs.

The ATT on prenatal visits in column 3 and delivery at a health facility in column 6 is positive and statistically significant.[21] The coefficient in column 3 shows that the HSAA leads to about 0.85 higher prenatal visits during pregnancy. Similarly, the HSAA has increased the probability of delivering at a health facility by 16.9 percent. In addition, the ATT in columns 4 and 5 for iron supplement intake and a tetanus shot during pregnancy are insignificant. Although the coefficient associated with the total number of vaccinations is negative[22], the overall pooled input corroborates the significant positive effect of the HSAA. The ATT in column 9 indicates that the amendment has increased overall parental investments in children by about 0.23 SDs.

---

[20] Following MacKinnon and Webb (2018), we address the small number of clusters by clustering standard errors at the state level and employing a bootstrap method with 1,000 iterations. Notably, clustering standard errors at the Primary Sampling Unit (PSU) level, which represents the smallest available geographic regions within the National Family Health Survey (NFHS), yields nearly identical results.
[21] According to the NFHS, mothers without formal education and those in the lowest quintile of wealth are much less likely to give birth in a hospital (Vora et al., 2009). Therefore, the fact that HSAA has had a significant positive effect on institutional delivery is not surprising.

[Table 2 about here]

The fundamental assumption of our estimation approach is the pre-treatment parallel trend, which entails observing similar trajectories of the outcomes when comparing treated and non-treated groups before the onset of treatment. Following Borusyak et al. (2021), we perform pre-trend testing by estimating equation (3) on the untreated observations. We use lead equals eight and treatment horizon equals 15 and visually examine the dynamic ATTs in Figure 2. The first thing to note here is that the ATTs before the treatment (i.e., the leads) are close to zero and statistically insignificant for both height (panel-A) and weight (Panel-B). We also conduct the conventional individual and joint tests of significance of $\gamma_p = 0$. In both individual and joint significance tests, we fail to reject the null of the non-zero treatment effect (see Appendix A1). Both visual exercise and statistical tests suggest that the parallel trend assumption holds.

The second point to note is that the HSAA effects become more prominent as the treatment horizon expands. Height, a long-term health and nutritional status indicator, requires a few years for a statistically significant treatment effect. On the contrary, short-term health and nutritional status, such as weight, exhibit an immediate and statistically significant treatment effect. Overall, the dynamic treatment effect figure indicates that the HSAA has a long-term impact on child health.

[Figure 2 about here]

*B. Placebo Exercises*

---

[22] The negative ATT coefficient for vaccination is driven by the large negative effect on the third or higher-order birth cohorts, an issue we explore in section C.

We conduct a falsification exercise to examine the validity of our estimation approach. Specifically, we re-estimate (2) using only the sample of non-Hindus (i.e., Muslims). Since the HSAA affects only the Hindu population, this placebo exercise should produce no effects of the HSAA on any of the outcomes for the Muslim sample. Table 3 reports the results based on this placebo test. The associated estimates for HFA and WFA outcomes are statistically insignificant at the 5-percent significance level. Similarly, the ATT on prenatal and postnatal inputs (reported in columns 4-8) and pooled inputs (reported in column 9) approaches zero and are statistically insignificant at the 5-percent signifcant level. Therefore, this placebo exercise bolsters our primary estimation approach and shows that the technique does not produce spurious effects on child health outcomes.

[Table 3 about here]

### C. Heterogeneous Treatment Analysis

We turn our attention to the impact of the HSAA on the health of later-born children and daughters, investigating whether these groups show marked health advancements. The HSAA's impact on human capital may depend on the child's gender or birth order. For instance, a cultural preference for a robust eldest son could be significant, particularly for Hindu families, given the prominence of their patrilocal and patrilineal kinship systems. Furthermore, Hindu religious doctrines underscore ceremonies exclusive to male heirs (Arnold, Choe, and Roy, 1998).

In calculating the heterogeneous treatment effects, we employ a method adapted from Borusyak et al. (2021), focusing on specific variables of interest. Such a method necessitates modifications to formula (2) with supplementary weights. As an illustration, the treatment effect for girls is ascertained by multiplying the standard treatment effects by individual weights and then normalizing them by the total count of female participants in the sample. We apply the same method to estimate the treatment effect for boys. After obtaining estimates for boys and girls, we use a linear combination of the coefficients such that $\hat{\tau}_{delta} = \hat{\tau}_{girls} - \hat{\tau}_{boy}$ and conduct a hypothesis test against a zero null. Failing to reject the null would indicate the absence of statistically significant heterogeneity between boys and girls.

We report the results for the heterogeneous effects in Tables 4 through 6. Table 4 reports the ATTs at the child level, showing the impacts on the same outcomes as reported in Table 2. We report the ATT estimates for boys and girls: the HFA z-score estimates (in column 1), the WFA z-score estimates (column 2), and the pool inputs index estimates (column 9). The results reveal a positive and significant ATT for both boys and girls. The difference between the ATT for girls and boys for height is not statistically significant, implying that we do not detect differential downstream effects of the HSAA by gender. However, the difference between the ATT for girls and boys is statistically significant for the weight outcome, indicating more substantial HSAA impacts on girls' weight than boys. However, we cannot make the same claim for the HFA z-score and parental inputs, as the differences between genders for these two outcomes are significant.

In Table 5, we report the ATT of the HSAA by birth order. The table reports the ATTs for the HSAA on height (in column 1), weight (in column 2), and parental care (in column 9). The results show positive and significant ATTs for height, weight, and the pooled parental care inputs, but only for first-born and second-born children. The ATT on third or later-born children is nearly zero and statistically insignificant. Thus, the HSAA has no positive effect on higher-born children and may even harm their health due to decreased parental care.

To test for both gender and birth order effects, we estimate the ATTs for birth order interacted with girls (i.e., first-born girl, second-born girl, and third or higher-born girl). We report the results in Table 6. They exhibit a similar pattern as those reported in Table 4: the later-born girls have lower height and weight. Later-born girls receive significantly lower parental inputs. Again, this finding implies that the HSAA did not necessarily benefit daughters of higher birth order.

[Table 4 about here] [Table 5 about here] [Table 6 about here]

*D. Mechanisms*

Although we have presented considerable evidence on how the HSAA improves child health, we have not yet addressed the channels through which the HSAA likely improved health outcomes.

Several potential mechanisms can explain our findings. Modeling household decisions for human capital investments is a challenging task. At the core of various explanations is the non-unitary model of household bargaining and how changes in household bargaining influence subsequent parental investments into the human capital for children. A large body of evidence from developing countries supports the idea that households are not unitary entities and that bargaining power is crucial (Duflo, 2012). If we assume two parents in a non-unitary household model, the changes in the HSAA inheritance law positively influence the female bargaining power within the household. The HSAA allows women to inherit property.

The HSAA can create multiple, theoretically coherent mechanisms for better child health. We explore these primary mechanisms at the household level before delving into data analysis for empirical validation. We do so by revisiting equation (2) and re-estimating it with each proposed mechanism serving as an outcome variable.

The HSAA could potentially incentivize women to postpone fertility and delay childbearing. An increase in bargaining power within marital relationships, granted by the enhanced resource control facilitated by HSAA, aligns with women's tendencies towards later and fewer childbirths. As a result, the HSAA may serve as a significant factor in moderating or deferring fertility.

The second channel relates to an increase in unearned income by women. The HSAA raises women's unearned income due to enhanced land inheritance access. A noncooperative household bargaining framework posits that the higher income could translate to better control over the woman's income, resulting in improved gains for the female in the household from her work. Heath and Tan (2020) show empirical evidence that the HSAA increases the female labor supply. This increase in the wife's labor supply does not adversely impact the husband's labor supply, allowing for a higher total household income resulting from the HSAA. Additional household income may influence height through access to an improved diet quality or increased calorie intake. Even if increased food intake does not directly come from an increase in total household income, reducing household size via reductions in childbearing (noted earlier) can lead to a higher nutritional intake per person.

Third, the disease environment in early childhood can play a critical role in influencing child health. Child height is a function of net nutrition, the difference between food intake and energy loss due to physical activities and disease. In India, especially in rural parts, diarrheal disease, fevers, and respiratory infections are high. These can impose a nutritive tax on one's nutritional status in early childhood and, subsequently, on height. There is clear evidence that inadequate prenatal nutrition, which is quite prevalent in India, causes low birth weight and worsens child health. Finally, infections during early childhood may affect height because they can sap the energy required for physical growth. Both respiratory and gastrointestinal diseases can affect height (Victora, 1990).

Finally, higher female bargaining power within the household can improve child outcomes. Existing research shows that an increase in women's income or bargaining power within the household benefits children more than increases in men's income (Thomas, 1990, Lundberg, Pollak, and Wales, 1997). When women hold a higher fraction of the household income, the households spend more on children's clothing and food. As a result, children in such families receive more and better nutrition. There are potentially two hypotheses put forth that could explain this phenomenon. The first relates to women and men having different preferences (the so-called *preference hypothesis*). It posits that women are more concerned about the well-being of their children. According to this *preference hypothesis*, when women receive income transfers (versus men), the trade-off is between additional spending on children versus private consumption by men. A second possibility is the *specialization hypothesis,* which could account for differential spending. According to this framework, women and men specialize in different tasks based on their comparative advantage. For example, women with lower wages than men specialize in time-intensive tasks, including childcare and food preparation. On the other hand, men will take charge of jobs that require money but little time, such as saving and investing.

*E. Supporting Evidence*

Data limitations prevent us from taking on one particular channel against the others—fulfilling the assumptions for formal mediation analysis is challenging in our context.[23] In what follows, we attempt to shed more light on the plausibility of some of the hypothesized channels using the available NFHS data. Table 7 reports the results.

[Table 7 about here]

**Childbearing**: In columns (1) and (2), we explore the childbearing channel with two outcomes: the total number of children *ever born* and the age at first birth. The results reported in Table 7 show evidence consistent with the idea that the total number of children born decreased among Hindu women living in HSAA-implementing states after the policy implementation (in column 1).[24] However, women's age at her first birth declined slightly by less than a month (in column 2), which is statistically insignificant at the 5-percent significance level and economically close to zero. This fertility decline reduces household size via lower childbearing, likely resulting in higher nutritional intake per person.

**Household Wealth:** The NFHS does not collect income measures or any specific proxies of earnings by a household member. Therefore, we use two survey questions to proxy household wealth to capture the potential income effects. We rely on survey questions, whether the household has electricity and whether the home has piped water, to proxy the change in household wealth in HSAA-implementing areas.[25] Columns (3) and (4) in Table 7 report the results. The results show that both ATTs are positive, and ATT for electricity is statistically significant. These are both consistent with a positive income effect.

---

[23] Conducting formal mediation analysis in our study is difficult due to the likely interdependence among mediators. Therefore, we emphasize the impact of the HSAA on key mediators without quantifying the extent to which a mediator explains the treatment effect.

[24] We argue that our sample's decline in fertility does not lead to differential childbearing. In India, there exists a strong first-born son preference (Jayachandran and Pande, 2017). Suppose son preference led to sex selective feticide and abortion. In that case, the composition of our sample might be selected, i.e., children we observe are only of the sex that the parents desire. Moreover, declining fertility may worsen sex-selective abortion. In that case, the positive effect is not because of the HSAA but rather because of sex-selective childbearing. However, we do not find any evidence of differential childbearing, see Appendix Table A2.

[25] Both access to electricity and piped drinking water are integral components of numerous wealth indices, such as the Demographic Health Survey's wealth index and the international wealth index (Rutstein and Johnson. 2004). Furthermore, empirical research indicates a positive correlation between access to electricity and piped water and increased wealth levels in developing nations (Ferguson, Wilkinson, and Hill, 2000 ).

**Disease Environment**: In column (5) of Table 7, we explore the possibility that the HSAA directly affects the household disease environment. We show the results on a composite index based on the following variables: whether the child had a fever in the last two weeks, the incidence of cough during the previous two weeks, and diarrhea in the last two weeks. The ATT of the disease environment index in column (5) is statistically significant. Therefore, empirical evidence supports that the positive height and weight impacts are likely to operate through a disease environment channel.

**Female Bargaining Power:** In Appendix A, Table A3, we show the HSAA improved marriage market outcomes for women living in treated areas. Women living in HSAA-implementing areas exhibit considerably lower age gaps than their husbands (see column 1). They are likelier to marry better-educated men (columns 2-3) and more wealthy husbands (column 4). In addition, we show that treated women report having similar fertility preferences as their husbands (column 5), and their husbands' families receive lower dowry payments[26] (column 6).

Furthermore, consistent with the primary hypothesized mechanism (bargaining power), treated women report a higher likelihood of receiving a land inheritance and owning more land acreage. Better marriage market outcomes and greater land ownership will likely increase women's bargaining power in the household.

We further investigate the women's bargaining channel using various empowerment measures available in the NFHS (see Table 7). Specifically, we use three decision measures: whether the woman has a say about large purchases (column 6), whether she has a voice in her own healthcare choices (column 7), and whether she goes to the market alone (column 8). We combine these measures into a composite female bargaining measure (column 9). The coefficients associated with the ATT for the female bargaining index are positive and statistically significant, implying that the female bargaining power increased due to the HSAA. Table A4, however, shows that women's bargaining power improved primarily among the top two asset quintile households.

---

[26] Notably, our estimated coefficient on dowry payments is negative, which stands in contrast to the positive estimate found in Roy (2015). We can reconcile this discrepancy because our estimation approach relies on the DiD imputation method by Borusyak et al. (2021), whereas Roy (2015) relies on the classical TWFE estimation approach. Roth et al. (2023) argue that the TWFE approach is likely to encounter a negative weighing issue and result in the opposite sign of treatment effect.

The above explanations are not necessarily mutually exclusive. However, we cannot rule out the possibility that some of these channels, in tandem, improved child health.

## F. Robustness Check

To explore the robustness of our results, we next employ the conventional difference-in-differences strategy (i.e., two-way fixed effect (TWFE) estimation). We estimate the effect of the HSAA using the following reduced-form specification:

$$H_{ist} = \beta_0 + \beta_1(Treat_s \times After_{st}) + X_{ist}\Pi + \alpha_s + \mu_t + \delta_{s,t} + \epsilon_{ist} \tag{4}$$

$H_{ist}$ is the outcome for child $i$ born to a mother who married in year $t$, in state $s$. $Treat$ is an indicator variable: it equals one if the mother is from a state that amended the HSA before the national adoption in 2005 (i.e., reform states) and is set to zero otherwise. $After_{st}$ is a dummy variable: it takes a value of one if the mother was married after the reform in her state $s$ and takes zero otherwise. $Treat \times After$ is a binary indicator, set to one if the mother is from a reform state $s$ and was married after the reform in her state $s$, and equals zero if she is unexposed to the reform. Finally, $\alpha_s$ is state fixed effect, $\mu_t$ is the mother's year of birth fixed effect[27], and $\delta_{s,t}$ captures state-year fixed effects. State fixed effects capture state-specific characteristics. Year fixed effects account for time-varying but group-invariant factors. The state-year fixed effect allows us to control state-specific time-varying omitted variables, which may correlate with the HSAA. $X_i$ is a set of mother-related characteristics (age at first birth, age at first birth squared, age); it also includes child characteristics, such as gender, age, and birth order.

The parameter of interest, $\beta_1$, captures the effect of the HSAA. The validity of our empirical approach depends on two assumptions: (a) no pre-trends exist for the treatment and control groups, and (b) states do not adopt the HSAA in a manner correlated with child health.

---

[27] We use mothers' year of birth fixed effects to control for maternal age, as several studies have demonstrated a relationship between maternal age and the risks of child mortality and undernutrition (Finlay, Ozaltin, and Canning, 2011).

We test whether there is a similar evolution pattern of child health in the treated and non-treated states before the reform using a regression approach. This exercise uses data from the treatment and control states before the reform. Specifically, we use data until 1984 as the first state amended the HSA in 1985. We estimate:

$$H_{ist} = \delta_0 + \delta_1(Treat_s) + \delta_2(Year_t) + \delta_3(Treat_s \times Year_t) + X_{ist}\Pi + \epsilon_{ist} \qquad (5)$$

where $H_{ist}$ is the outcome for child $i$ born to a mother who married in year $t$, in state $s$. $Treat$ is a dummy variable that equals one if the mother is from a state that amended HSA before the national adoption in 2005 (i.e., reform states) and zero if not from one of the reform states. $Year_t$ is a linear trend of the mother's year of marriage from 1970 to 1984. $X_i$ is the same set of controls as in equation (1). $\delta_3$ captures the differential trend in the outcome variable between treated and non-treated states before the HSAA.

We report the results of this exercise in Table A5. The coefficients of $Treat \times Year$ in columns 1 and 3 are insignificant and close to zero for HFA and WFA. Including controls (reported in columns 2 and 4) only strengthens these results. Therefore, this exercise further bolsters the validity of the empirical approach.

We formally investigate the possible selection into the HSAA by testing the correlation between the HSAA and average child health with data before the amendments. The earliest survey years available for child health from the NFHS-1 are during the 1992-1993 period. We aim to examine the relationship between the HSAA and health outcomes during the baseline period. We focus on the states that adopted the HSAA after 1993: Maharashtra (1994) and Karnataka (1994). Unlike the subsequent survey waves used in the primary analysis, data in this particular wave is also available at the district level. Therefore, we analyze the health outcomes for Hindu-only households at the district level. We estimate the following equation:

$$HSAA_d = \gamma_0 + \gamma_1 H_d^{1993} + X_d^{1993}\Pi + \eta_d \qquad (6)$$

$HSAA_d$ is a binary indicator, and it is set to one if the district is subject to the HSA amendment after 1993 and before the national adoption in 2005. This indicator is set to zero otherwise; $H_d^{1993}$ is the child health outcomes–HFA and WFA z-scores– in district $d$ in the year 1992-93; $X_d^{1993}$ is the vector of controls capturing the average of various socioeconomic characteristics for the district in 1992-93 (i.e., mother's age at first birth, mother's age at first birth squared, mother's current age dummies). We estimate equation (6) twice for the HFA and WFA outcomes. Table A6 reports the results.

We first regress the district's HSAA status on the average HFA z-score in 1992-93 without other controls (column 1 of Table A6 reports the results). The coefficient of HFA is close to zero and statistically insignificant. Including controls in column 2 does not alter the result, and we find no correlation between HSAA adoption and HFA outcomes. We obtain similar results for WFA outcomes (at the district level), thereby bolstering our claim regarding the absence of a relationship between the HSAA and child health outcomes.

Table A7 reports the estimates of equation (4). The dependent variables are the primary health outcomes and the parental investment in child health. Our variable of interest, $Treat \times After$, is a dummy capturing whether the mother is from a reform state $s$ and was married after the reform. The table contains the OLS coefficients and standard errors in parentheses. We cluster the standard errors at the primary sampling unit (PSU). All columns of Table A7 include control variable and fixed effects (FE)– state fixed effect, year (mother's year of birth) fixed effect, state-year fixed effect, and survey year fixed effect.

The results reported in Table A7 suggest that India's women-favored inheritance rights reform has significantly improved child height and weight. The estimates of the health outcomes– height-for-age (HFA) z-score and weight-for-age (WFA) z-score– are presented in columns 1 and 2. The coefficients of $Treat \times After$ in columns 1 and 2 are positive and statistically significant. For example, in column 1, the coefficient of $Treat \times After$ is 0.183, which indicates that the HSAA has, on average, increased children's HFA z-score by 0.183 SDs. Similarly, the coefficient of $Treat \times After$ in column 2 indicates that the HSAA has, on average increased children's WFA z-score by 0.276 SDs. The coefficient of $Treat \times After$ in column 9 indicates that the amendment has increased overall parental investment in children by about 0.36 SDs. These estimates are similar to our baseline results reported in Table 2. We also present the heterogeneity by child's sex, birth order, and child's sex interacted with birth order in Appendix A8 through Table A10.

# VI. Conclusion

Child height is a significant predictor of adult human capital and economic outcomes. Despite this fact, India faces a stunting rate of 31 percent, which is remarkably high even among developing nations. In this study, we investigate the impact of inheritance law amendments, specifically the Hindu Succession Act (HSA) changes, on child height in India. These amendments sought to provide unmarried women with equal shares of ancestral property, potentially transforming their inheritance rights.

Our findings reveal convincing evidence that implementing these amendments significantly improved child health. However, these health benefits mainly apply to first-born and second-born children, with no substantial improvements observed for later-born daughters. To support the role of enhanced parental care, we demonstrate increased prenatal and postnatal investments in regions that implemented the amendments. We link the outcomes to four factors: improved female bargaining power, enhanced home disease environment, fertility choices, and increased household net wealth.

Our findings underscore the positive impact of enhanced female autonomy on child health at the household level. In addition, they align with prior research suggesting a consistent improvement in outcomes within women's sphere of influence, particularly when they command greater control over their income. In developing countries like India, human capital development offers a potential escape from poverty. Our findings contribute to a broader understanding of how legal reforms, specifically those addressing gender inequality, can have far-reaching implications for child health and overall family well-being.

Our study implies that enhancing women's inheritance rights may yield long-term consequences beyond economic well-being, impacting their children's human capital accumulation. Policymakers should consider these results when planning future reforms that aim to promote gender equity and improve the lives of children in developing countries. However, it is crucial to acknowledge that the positive effects observed in this study do not apply evenly to all children. This finding highlights the need for additional measures and interventions to address health disparities among later-born and female children.

# Figures and Tables

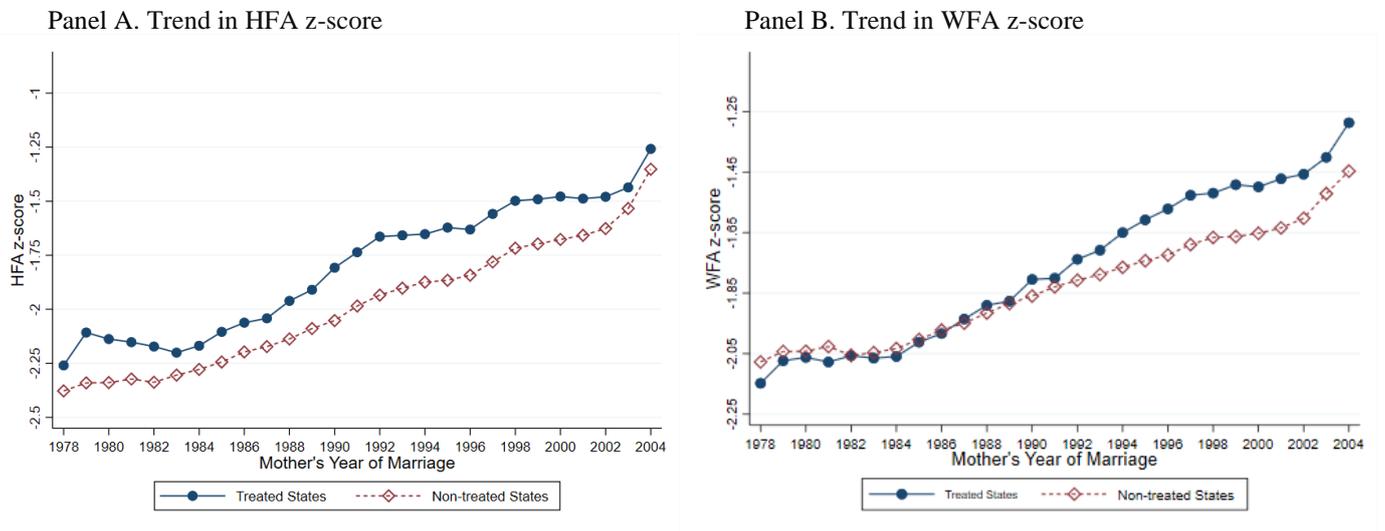

**Figure 1:** Trends in Child Health Outcomes in Treated and Non-Treated Areas.

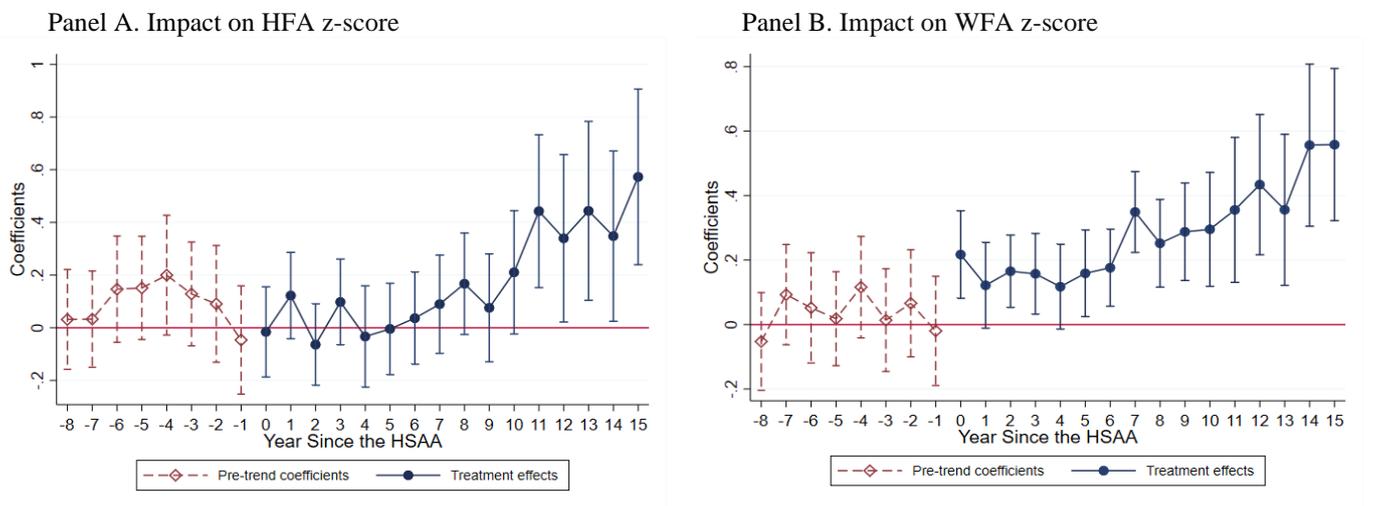

**Figure 2:** The Impact of HSAA on Child Health Outcomes

**Table 1**: Summary Statistics

|  | Full Sample | Treated States | | | Non-Treated States | | |
|---|---|---|---|---|---|---|---|
|  |  | Treated Religion | Non-Treated Religion | All | Treated Religion | Non-Treated Religion | All |
|  | (1) | (2) | (3) | (4) | (5) | (6) | (7) |
| Household Has Electricity (1=if yes) | 0.629 | 0.823 | 0.907 | 0.835 | 0.578 | 0.545 | 0.575 |
|  | (0.482) | (.382) | (0.294) | (0.371) | (0.494) | (0.498) | (0.494) |
| Household Receives Piped Water at Home (=1 if yes) | 0.309 | 0.504 | 0.686 | 0.531 | 0.247 | 0.275 | 0.250 |
|  | (0.462) | (0.50) | (0.464) | (0.499) | (0.431) | (0.447) | (0.433) |
| Mother's Age (Years) | 26.05 | 24.81 | 25.58 | 24.93 | 26.22 | 27.28 | 26.35 |
|  | (5.37) | (4.72) | (5.28) | (4.814) | (5.38) | (6.02) | (5.47) |
| Mother's Age at First Marriage (Years) | 17.27 | 17.60 | 17.39 | 17.57 | 17.24 | 16.86 | 17.20 |
|  | (3.20) | (3.51) | (3.10) | (3.45) | (3.16) | (2.78) | (3.12) |
| Total Children Born to a Mother | 3.03 | 2.383 | 3.01 | 2.476 | 3.081 | 3.942 | 3.183 |
|  | (1.91) | (1.37) | (1.83) | (1.467) | (1.904) | (2.385) | (1.986) |
| Say About a Large Purchase (=1 if yes) | 0.490 | 0.515 | 0.482 | 0.510 | 0.480 | 0.505 | 0.483 |
|  | (0.500) | (0.500) | (0.500) | (0.500) | (0.500) | (0.500) | (0.500) |
| Say About Own Health (=1 if yes) | 0.545 | 0.569 | 0.557 | 0.568 | 0.531 | 0.585 | 0.538 |
|  | (0.498) | (0.495) | (0.497) | (0.495) | (0.499) | (0.493) | (0.499) |
| Go to Market Alone (=1 if yes) | 0.400 | 0.548 | 0.431 | 0.531 | 0.361 | 0.350 | 0.360 |
|  | (0.490) | (0.498) | (0.495) | (0.499) | (0.480) | (0.477) | (0.480) |
| Bargaining | 0.000 | 0.153 | 0.027 | 0.135 | -0.050 | 0.024 | -0.041 |
|  | (1.00) | (0.991) | (1.011) | (0.995) | (0.997) | (0.999) | (0.998) |
| Child is a Girl (=1 if yes) | 0.480 | 0.476 | 0.491 | 0.478 | 0.479 | 0.492 | 0.480 |
|  | (0.500) | (0.499) | (0.500) | (0.500) | (0.500) | (0.500) | (0.500) |
| Child's Birth Order | 2.805 | 2.182 | 1.729 | 2.263 | 2.855 | 3.656 | 2.945 |
|  | (1.897) | (1.362) | (1.840) | (1.456) | (1.892) | (2.379) | (1.973) |
| Child's HFA z-score | -1.935 | -1.665 | -1.675 | -1.666 | -1.985 | -2.163 | -2.006 |
|  | (1.850) | (1.737) | (1.735) | (1.736) | (1.86) | (1.928) | (1.869) |
| Child's WFA z-score | -1.805 | -1.663 | -1.555 | -1.647 | -1.835 | -1.939 | -1.847 |
|  | (1.374) | (1.293) | (1.313) | (1.297) | (1.391) | (1.380) | (1.390) |
| Childhood Disease Environment Index | 0.000 | -0.058 | -0.034 | -0.054 | 0.006 | 0.081 | 0.014 |
|  | (1.00) | (0.946) | (1.010) | (0.956) | (1.003) | (1.061) | (1.011) |
| Number of Children | 65,371 | 11,680 | 2,036 | 13,716 | 45,581 | 6,074 | 51,655 |

*Notes:* (a) Standard deviation appears in parentheses. (b) Treatment refers to Hindu Succession Act (HSA) 1956 amendment. Treated states are states that amended the HSA 1956. Treated religion refers to Hindus (i.e., Hindu, Jain, Sikh, and Buddhist) subject to the HSAA; non-treated religion refers to Muslims. (c) The entire sample in column 1 includes both treated and non-treated observations. All in Column 4 refers to observations in the treated states, and All in Column 7 refers to observations in the non-treated states. (d) The following variables are summarized at the mother level: mother's age (years), mother's age at first marriage (years), say about a large purchase (=1 if yes), say about own health (=1 if yes), go to market alone (=1 if yes), and bargaining. (e) Say about a large purchase takes a value of 1 if the mother has some say in household significant purchase decisions and zero otherwise. Say about own health takes a value 1 if the mother has some say in health care choices, and zero otherwise. Go to market alone takes a value of 1 if the mother can travel to the market alone, and zero otherwise. (f) Bargaining is defined as the first principal component of the three decision variables (i.e., say about a large purchase, say about own health, and go to market alone), and normalized it to have zero mean and one standard deviation. (g) HFA z-score is the child's height-for-age z-score, and WFA z-score is the child's weight-for-age z score. These z-scores are created by comparing the sample with the WHO growth standards. (h) The childhood disease environment index is the first principal component of the three variables (the child had a fever, cough, and diarrhea in the last two weeks) and normalized to mean zero and standard deviation one.

**Table 2**: Effect of HSAA on Child Health

| | HFA z-score | WFA z-score | Prenatal Inputs | | | Postnatal Inputs | | | |
| | | | Total Prenatal Visits During Pregnancy | Mother Took Iron Supplement During Pregnancy (1= if yes) | Mother Took Tetanus Shot During Pregnancy (=1 if yes) | Delivery at Health Facility (1= if yes) | Postnatal Check Within Two Months of Birth (1= if yes) | Total Vaccination | Pooled Inputs |
|---|---|---|---|---|---|---|---|---|---|
| | (1) | (2) | (3) | (4) | (5) | (6) | (7) | (8) | (9) |
| ATT | 0.144** | 0.271*** | 0.877*** | -0.024 | 0.005 | 0.169*** | 0.033 | -0.410*** | 0.234*** |
| | (0.067) | (0.049) | (0.116) | (0.017) | (0.014) | (0.017) | (0.022) | (0.102) | (0.041) |
| State FE | Yes | Yes | Yes | Yes | Yes | Yes | Yes | Yes | Yes |
| Year FE | Yes | Yes | Yes | Yes | Yes | Yes | Yes | Yes | Yes |
| Controls | Yes | Yes | Yes | Yes | Yes | Yes | Yes | Yes | Yes |
| Observations | 57,260 | 56,904 | 50,739 | 33,859 | 33,897 | 57,224 | 24,116 | 41,451 | 23,999 |

*Note:* (a) This table reports the average treatment effect on the treated (ATT) of HSAA on child health outcomes. (b) A woman is treated if she is from a state that amended HSA and married after the amendment in her state. (c) HFA z-score is the child's height-for-age z-score, and WFA z-score is the child's weight-for-age z score. (d) Outcome variables in columns 3-6 are prenatal inputs, and Columns 7-8 are postnatal inputs. (e) The mother took the iron supplement, and tetanus shots during pregnancy are available only for the most recent birth (youngest child). (f) Delivery at a health facility refers to childbirth at a health facility. (g) Total vaccination uses children ages 13–59 months, as the recommended age for some is up to one year. (h) Postnatal check within two months of birth is available only for the most recent birth (youngest child) and only in NFHS-2 and NHFS-3. (i) The outcome variable in column 9 is pooled inputs: Total prenatal visits during pregnancy, mother took iron supplement during pregnancy, mother took tetanus shot during pregnancy, delivery at a health facility, postnatal check within two months of birth, and child ever vaccinated. We take the first principal component of the standardized inputs and then normalized it to mean zero and standard deviation. (j) Bootstrap (1,000 iterations) standard errors are clustered by state and appear in parentheses. (k) State FE and Year FE (women's birth year) are added in all columns. (l) Control variables are the child's sex, child age (in months), child's birth order, household size (number of household members), and survey year. (l) Observations used in this table are Hindus (i.e., Hindus, Buddhists, Jains, and Sikhs) only as they are subject to HSA 1956. (m) ***Significant at the 1 percent level, **Significant at the 5 percent level, and *Significant at the 10 percent level.

**Table 3**: Placebo Treatment Effect of HSAA on Child Health

| | HFA z-score | WFA z-score | Prenatal Inputs | | | | Postnatal Inputs | | Pooled Inputs |
| --- | --- | --- | --- | --- | --- | --- | --- | --- | --- |
| | | | Total Prenatal Visits During Pregnancy | Mother Took Iron Supplement During Pregnancy (1= if yes) | Mother Took Tetanus Shot During Pregnancy (=1 if yes) | Delivery at Health Facility (1= if yes) | Postnatal Check Within Two Months of Birth (1= if yes) | Total Vaccination | |
| | (1) | (2) | (3) | (4) | (5) | (6) | (7) | (8) | (9) |
| ATT | -0.172 | 0.122 | 0.381 | -0.049 | -0.049 | -0.000 | -0.084 | 0.050 | -0.118 |
| | (0.182) | (0.116) | (0.246) | (0.052) | (0.040) | (0.042) | (0.050) | (0.364) | (0.086) |
| State FE | Yes | Yes | Yes | Yes | Yes | Yes | Yes | Yes | Yes |
| Year FE | Yes | Yes | Yes | Yes | Yes | Yes | Yes | Yes | Yes |
| Controls | Yes | Yes | Yes | Yes | Yes | Yes | Yes | Yes | Yes |
| Observations | 1,449 | 1,441 | 1,273 | 830 | 836 | 1,448 | 451 | 1,448 | 450 |

*Note*: (a) This table reports the average treatment effect on the treated (ATT) of HSAA on child health outcomes. (b) A woman is treated if she is from a state that amended HSA and married after the amendment in her state. (c) HFA z-score is the child's height-for-age z-score, and WFA z-score is the child's weight-for-age z score. (d) Outcome variables in columns 3-6 are prenatal inputs, and Columns 7-8 are postnatal inputs. (e) The mother took the iron supplement, and tetanus shots during pregnancy are available only for the most recent birth (youngest child). (f) Delivery at a health facility refers to childbirth at a health facility. (g) Total vaccination uses children ages 13–59 months, as the recommended age for some is up to one year. (h) Postnatal check within two months of birth is available only for the most recent birth (youngest child) and only in NFHS-2 and NHFS-3. (i) The outcome variable in column 9 is pooled inputs: total prenatal visits during pregnancy, mother took iron supplement during pregnancy, mother took tetanus shot during pregnancy, delivery at a health facility, postnatal check within two months of birth, and child ever vaccinated. We take the first principal component of the standardized inputs and then normalized it to mean zero and standard deviation one. (j) Bootstrap (1,000 iterations) standard errors are clustered by state and appear in parentheses. (k) State FE and Year FE (women's birth year) are added in all columns. (l) Control variables are the child's sex, child age (in months), child's birth order, household size (number of household members), and survey year. (l) Observations in this table are Muslims not subject to HSA 1956. (m) ***Significant at the 1 percent level, **Significant at the 5 percent level, and *Significant at the 10 percent level.

## Table 4: Heterogeneous Treatment Effect of HSAA on Child Health by Child's Sex

| | | | Prenatal Inputs | | | Postnatal Inputs | | | |
|---|---|---|---|---|---|---|---|---|---|
| | HFA z-score | WFA z-score | Total Prenatal Visits During Pregnancy | Mother Took Iron Supplement During Pregnancy (1= if yes) | Mother Took Tetanus Shot During Pregnancy (=1 if yes) | Delivery at Health Facility (1= if yes) | Postnatal Check Within Two Months of Birth (1= if yes) | Total Vaccination | Pooled Inputs |
| | (1) | (2) | (3) | (4) | (5) | (6) | (7) | (8) | (9) |
| ATT Girls | 0.166** | 0.310*** | 0.897*** | -0.014 | 0.016 | 0.170*** | 0.028 | -0.246** | 0.257*** |
| | (0.072) | (0.052) | (0.133) | (0.018) | (0.016) | (0.019) | (0.025) | (0.117) | (0.044) |
| ATT Boys | 0.129* | 0.239*** | 0.855*** | -0.032* | -0.003 | 0.167*** | 0.037 | -0.552*** | 0.209*** |
| | (0.072) | (0.050) | (0.131) | (0.017) | (0.015) | (0.019) | (0.024) | (0.116) | (0.044) |
| ATT Girls - ATT Boys | 0.038 | 0.071** | 0.042 | 0.018* | 0.019** | 0.003 | 0.009 | 0.306*** | 0.048 |
| | (0.042) | (0.031) | (0.084) | (0.010) | (0.008) | (0.010) | (0.016) | (0.058) | (0.031) |
| State FE | Yes | Yes | Yes | Yes | Yes | Yes | Yes | Yes | Yes |
| Year FE | Yes | Yes | Yes | Yes | Yes | Yes | Yes | Yes | Yes |
| Observations | 57,260 | 56,904 | 50,739 | 33,859 | 33,897 | 57,224 | 24,116 | 41,451 | 23,999 |

*Note:* (a) This table reports the average treatment effect on the treated (ATT) of HSAA on child health outcomes. (b) A woman is treated if she is from a state that amended HSA and married after the amendment in her state. (c) HFA z-score is the child's height-for-age z-score, and WFA z-score is the child's weight-for-age z score. (d) Outcome variables in columns 3-6 are prenatal inputs, and Columns 7-8 are postnatal inputs. (e) The mother took the iron supplement, and tetanus shots during pregnancy are available only for the most recent birth (youngest child). (f) Delivery at a health facility refers to childbirth at a health facility. (g) Total vaccination uses children ages 13–59 months, as the recommended age for some is up to one year. (h) Postnatal check within two months of birth is available only for the most recent birth (youngest child) and only in NFHS-2 and NHFS-3. (i) The outcome variable in column 9 is pooled inputs: total prenatal visits during pregnancy, mother took iron supplement during pregnancy, mother took tetanus shot during pregnancy, delivery at a health facility, postnatal check within two months of birth, and child ever vaccinated. We take the first principal component of the standardized inputs and then normalized it to mean zero and standard deviation one. (j) Bootstrap (1,000 iterations) standard errors are clustered by state and appear in parentheses. (k) State FE and Year FE (women's birth year) are added in all columns. (l) Control variables are the child's sex, child age (in months), child's birth order, household size (number of household members), and survey year. (m) Observations used in this table are Hindus (i.e., Hindus, Buddhists, Jains, and Sikhs) only as they are subject to HSA 1956. (n) ***Significant at the 1 percent level, **Significant at the 5 percent level, and *Significant at the 10 percent level.

**Table 5**: Heterogeneous Treatment Effect of HSAA on Child Health by Child's Birth Order

| | HFA z-score | WFA z-score | Total Prenatal Visits During Pregnancy | Mother Took Iron Supplement During Pregnancy (1= if yes) | Mother Took Tetanus Shot During Pregnancy (=1 if yes) | Delivery at Health Facility (1= if yes) | Postnatal Check Within Two Months of Birth (1= if yes) | Total Vaccination | Pooled Inputs |
|---|---|---|---|---|---|---|---|---|---|
| | | | Prenatal Inputs | | | | Postnatal Inputs | | |
| | (1) | (2) | (3) | (4) | (5) | (6) | (7) | (8) | (9) |
| ATT 1st Child | 0.279*** | 0.406*** | 1.619*** | 0.029 | 0.036** | 0.273*** | 0.120*** | -0.080 | 0.472*** |
| | (0.084) | (0.062) | (0.172) | (0.020) | (0.018) | (0.025) | (0.027) | (0.140) | (0.060) |
| ATT 2nd Child | 0.211** | 0.317*** | 1.010*** | -0.011 | 0.030 | 0.181*** | 0.006 | -0.271* | 0.198*** |
| | (0.096) | (0.07) | (0.184) | (0.023) | (0.021) | (0.027) | (0.034) | (0.163) | (0.071) |
| ATT 3rd+ Child | -0.029 | 0.094 | -0.157 | -0.082** | -0.028 | 0.037 | -0.107* | -0.708** | -0.150 |
| | (0.171) | (0.131) | (0.362) | (0.038) | (0.033) | (0.055) | (0.056) | (0.327) | (0.134) |
| ATT 1st Child - ATT 2nd Child | 0.067 | 0.089 | 0.610*** | 0.040* | 0.006 | 0.092*** | 0.114*** | 0.192 | 0.273*** |
| | (0.097) | (0.074) | (0.195) | (0.022) | (0.018) | (0.026) | (0.035) | (0.169) | (0.076) |
| ATT 1st Child - ATT 3rd+ Child | 0.308* | 0.312** | 1.777*** | 0.111*** | 0.065* | 0.237*** | 0.227*** | 0.628* | 0.622*** |
| | (0.188) | (0.146) | (0.422) | (0.042) | (0.034) | (0.065) | (0.060) | (0.377) | (0.156) |
| State FE | Yes | Yes | Yes | Yes | Yes | Yes | Yes | Yes | Yes |
| Year FE | Yes | Yes | Yes | Yes | Yes | Yes | Yes | Yes | Yes |
| Observations | 57,260 | 56,904 | 50,739 | 33,859 | 33,897 | 57,224 | 24,116 | 41,451 | 23,999 |

*Note:* (a) This table reports the average treatment effect on the treated (ATT) of HSAA on child health outcomes. (b) A woman is treated if she is from a state that amended HSA and married after the amendment in her state. (c) The HFA z-score is the child's height-for-age z-score, and the WFA z-score is the child's weight-for-age z score. (d) Outcome variables in columns 3-6 are prenatal inputs, and Columns 7-8 are postnatal inputs. (e) The mother took the iron supplement, and tetanus shots during pregnancy are available only for the most recent birth (youngest child). (f) Delivery at a health facility refers to childbirth at a health facility. (g) Total vaccination uses children ages 13–59 months, as the recommended age for some is up to one year. (h) Postnatal check within two months of birth is available only for the most recent birth (youngest child) and only in NFHS-2 and NHFS-3. (i) The outcome variable in column 9 is pooled inputs: total prenatal visits during pregnancy, mother took iron supplement during pregnancy, mother took tetanus shot during pregnancy, delivery at a health facility, postnatal check within two months of birth, and child ever vaccinated. We take the first principal component of the standardized inputs and then normalized it to mean zero and standard deviation one. (j) Bootstrap (1,000 iterations) standard errors are clustered by state and appear in parentheses. (k) State FE and Year FE (women's birth year) are added in all columns. (l) Control variables are the child's sex, child age (in months), child's birth order, household size (number of household members), and survey year. (m) Observations used in this table are Hindus (i.e., Hindus, Buddhists, Jains, and Sikhs) only as they are subject to HSA 1956. (n) ***Significant at the 1 percent level, **Significant at the 5 percent level, and *Significant at the 10 percent level.

**Table 6**: Heterogeneous Treatment Effect of HSAA on Child Health by Girl Child at Different Birth Orders

| | | | Prenatal Inputs | | | Postnatal Inputs | | | |
|---|---|---|---|---|---|---|---|---|---|
| | HFA z-score | WFA z-score | Total Prenatal Visits During Pregnancy | Mother Took Iron Supplement During Pregnancy (1= if yes) | Mother Took Tetanus Shot During Pregnancy (=1 if yes) | Delivery at Health Facility (1= if yes) | Postnatal Check Within Two Months of Birth (1= if yes) | Total Vaccination | Pooled Inputs |
| | (1) | (2) | (3) | (4) | (5) | (6) | (7) | (8) | (9) |
| ATT 1st Child Girl | 0.307*** | 0.458*** | 1.645*** | 0.029 | 0.048** | 0.276*** | 0.108*** | 0.063 | 0.498*** |
| | (0.102) | (0.077) | (0.214) | (0.026) | (0.022) | (0.030) | (0.034) | (0.178) | (0.075) |
| ATT 2nd Child Girl | 0.211* | 0.338*** | 0.994*** | 0.009 | 0.045* | 0.179*** | 0.008 | -0.060 | 0.233** |
| | (0.120) | (0.091) | (0.229) | (0.028) | (0.025) | (0.033) | (0.043) | (0.215) | (0.091) |
| ATT 3rd+ Child Girl | 0.019 | 0.128 | -0.109 | -0.066 | -0.028 | 0.041 | -0.110 | -0.576 | -0.149 |
| | (0.250) | (0.192) | (0.5491) | (0.057) | (0.048) | (0.071) | (0.078) | (0.497) | (0.188) |
| ATT 1st Child Girl - ATT 2nd Child Girl | 0.094 | 0.119 | 0.654** | 0.020 | 0.003 | 0.097** | 0.100** | 0.123 | 0.265** |
| | (0.141) | (0.110) | (0.290) | (0.033) | (0.026) | (0.038) | (0.050) | (0.259) | (0.109) |
| ATT 1st Child Girl - ATT 3rd+ Child Girl | 0.287 | 0.329 | 1.756*** | 0.094 | 0.076 | 0.235*** | 0.218** | 0.639 | 0.647*** |
| | (0.287) | (0.222) | (0.585) | (0.066) | (0.054) | (0.084) | (0.090) | (0.582) | (0.222) |
| State FE | Yes | Yes | Yes | Yes | Yes | Yes | Yes | Yes | Yes |
| Year FE | Yes | Yes | Yes | Yes | Yes | Yes | Yes | Yes | Yes |
| Observations | 57,260 | 56,904 | 50,739 | 33,859 | 33,897 | 57,224 | 24,116 | 41,451 | 23,999 |

*Note:* (a) This table reports the average treatment effect on the treated (ATT) of HSAA on child health outcomes. (b) A woman is treated if she is from a state that amended HSA and married after the amendment in her state. (c) The HFA z-score is the child's height-for-age z-score, and the WFA z-score is the child's weight-for-age z score. (d) Outcome variables in columns 3-6 are prenatal inputs, and Columns 7-8 are postnatal inputs. (e) The mother took the iron supplement, and tetanus shots during pregnancy are available only for the most recent birth (youngest child). (f) Delivery at a health facility refers to childbirth at a health facility. (g) Total vaccination uses children ages 13–59 months, as the recommended age for some is up to one year. (h) Postnatal check within two months of birth is available only for the most recent birth (youngest child) and only in NFHS-2 and NHFS-3. (i) The outcome variable in column 9 is pooled inputs: total prenatal visits during pregnancy, mother took iron supplement during pregnancy, mother took tetanus shot during pregnancy, delivery at a health facility, postnatal check within two months of birth, and child ever vaccinated. We take the first principal component of the standardized inputs and then normalize it to have a mean of zero and a standard deviation of one. (j) Bootstrap (1,000 iterations) standard errors are clustered by state and appear in parentheses. (k) State FE and Year FE (women's birth year) are added in all columns. (l) Control variables are the child's sex, child age (in months), child's birth order, household size (number of household members), and survey year. (m) Observations used in this table are Hindus (i.e., Hindus, Buddhists, Jains, and Sikhs) only as they are subject to HSA 1956. (n) ***Significant at the 1 percent level, **Significant at the 5 percent level, and *Significant at the 10 percent level.

**Table 7**: Mechanism of the Effect of HSAA on Child Health

| | Fertility Choices | | Household Wealth | | Disease Environment | Female Bargaining | | | |
|---|---|---|---|---|---|---|---|---|---|
| | Total Children Born | Age at First Birth (in Years) | Have Electricity (=1 if yes) | Receive Piped Water at Home (=1 if yes) | Disease Environment Index | Say About a Large Purchase (=1 if yes) | Say About Own Health (=1 if yes) | Going to the Market Alone (=1 if yes) | Bargaining Index |
| | (1) | (2) | (3) | (4) | (5) | (6) | (7) | (8) | (9) |
| ATT | -0.324*** | -0.072* | 0.092*** | 0.028 | -0.126*** | 0.093*** | 0.070*** | -0.002 | 0.164*** |
| | (0.085) | (0.041) | (0.020) | (0.024) | (0.040) | (0.022) | (0.022) | (0.020) | (0.044) |
| State FE | Yes | Yes | Yes | Yes | Yes | Yes | Yes | Yes | Yes |
| Year FE | Yes | Yes | Yes | Yes | Yes | Yes | Yes | Yes | Yes |
| Controls | Yes | Yes | Yes | Yes | Yes | Yes | Yes | Yes | Yes |
| Observations | 57,260 | 57,260 | 55,559 | 57,260 | 57,119 | 40,152 | 40,153 | 40,408 | 40,152 |

*Note:* (a) A woman is treated if she is from a state that amended HSA and married after the amendment in her state. (b) We use two indicators of women's fertility choices: the number of children born and the mother's age at first marriage. Total child born is the number of children born to a mother. (c) We use two indicators of household wealth: access to electricity and piped water. (d) The disease environment of the household is measured by the incidence of disease in the last two weeks. The child's disease in the last weeks is defined as the first principal component of the three variables (the child had a fever, cough, and diarrhea in the previous two weeks) and normalized to mean zero and standard deviation one. (e) We use three indicators of bargaining: the woman has a say in large purchases and her own health care and can go to market alone. Bargaining in column 9 is defined as the first principal component of the three decision variables (say about large purchases, say about own health, and going to market alone) and normalized to have zero mean and 1 standard deviation. (f) Bootstrap (1,000 iterations) standard errors are clustered by state and appear in parentheses. (g) State FE and Year FE (mother's birth year) are added in all columns. (h) The control variable is the survey year. (i) Observations used in this table are Hindus (i.e., Hindus, Buddhists, Jains, and Sikhs) only as only they are subject to HSA 1956. (j)***Significant at the 1 percent level, **Significant at the 5 percent level, *Significant at the 10 percent level.

# Online Appendix A

**Table A1**: Pre-trend Testing

|  | HFA z-score | WFA z-score |
|---|---|---|
| **Panel A: Individual Test of Significance** | | |
| Lead1 | -0.047 | -0.020 |
|  | (0.105) | (0.086) |
| Lead2 | 0.091 | 0.066 |
|  | (0.113) | (0.085) |
| Lead3 | 0.128 | 0.014 |
|  | (0.101) | (0.081) |
| Lead4 | 0.200 | 0.116 |
|  | (0.116) | (0.080) |
| Lead5 | 0.151 | 0.018 |
|  | (0.100) | (0.074) |
| Lead6 | 0.146 | 0.052 |
|  | (0.103) | (0.087) |
| Lead7 | 0.033 | 0.093 |
|  | (0.093) | (0.079) |
| Lead8 | 0.032 | -0.053 |
|  | (0.097) | (0.077) |
| **Panel B: Joint Test of Significance** | | |
| Null: Lead1 = Lead2 = . . . = Lead8 = 0 | | |
| Chi2 Value | 7.63 | 5.93 |
| Prob > Chi2 | 0.471 | 0.655 |

*Note:* (a) This table reports the average treatment effect on the treated (ATT) of HSAA on child health outcomes before the actual treatment happened. We present both individual tests of significance and joint tests of significance results. (b) HFA z-score is the child's height-for-age z-score, and WFA z-score is the child's weight-for-age z score. (c) Bootstrap (1,000 iterations) standard errors are clustered by state and appear in parentheses. (d) State FE and Year FE (women's birth year) are added in all columns. (e) Control variables are the child's sex, age (in months), birth order, household size (number of household members), and survey year. (f) Observations used in this table are Hindus (i.e., Hindus, Buddhists, Jains, and Sikhs) only, subject to HSA 1956. (g) ***Significant at the 1 percent level, **Significant at the 5 percent level, and *Significant at the 10 percent level.

**Table A2**: Effect of HSAA on Composition of Children

|  | Sex Composition of Children | | | |
|---|---|---|---|---|
|  | Sons Count | Son Proportion in Family | Have More Sons than Daughters (=1 if yes) | Have a Daughter (=1 if yes) |
|  | (1) | (2) | (3) | (4) |
| ATT | 0.006 | 0.004 | -0.017 | -0.018 |
|  | (0.043) | (0.019) | (0.019) | (0.019) |
| Observations | 57,260 | 57,260 | 57,260 | 57,260 |
| State FE | Yes | Yes | Yes | Yes |
| Year FE | Yes | Yes | Yes | Yes |
| Controls | Yes | Yes | Yes | Yes |

Note: (a) A woman is treated if she is from a state that amended HSA and married after the amendment in her state. (b) We use four indicators of the sex composition of women: the total number of sons, the proportion of children who are sons, have more sons than daughters, and have a daughter. (c) Bootstrap (1,000 iterations) standard errors are clustered by state and appear in parentheses. (d) State FE and Year FE (mother's birth year) are added in all columns. (e) Control variables are survey year. (f) The sample used in this table consists of Hindus (i.e., Hindus, Buddhists, Jains, and Sikhs) only as only they are subject to HSA 1956. (g)***Significant at the 1 percent level, **Significant at the 5 percent level, *Significant at the 10 percent level.

Table A3: Effect of HSAA on Women's Marriage Market Outcomes and Land Ownership

| | Marriage Market Outcomes | | | | | | Land Ownership | |
|---|---|---|---|---|---|---|---|---|
| | Age Gap | Husband's Education (Years) | Husband Has a College Degree (=1 if yes) | Top Wealth Quintile (=1 if yes) | Husband and Wife Desire the Same Fertility Outcome | Log (Cash Dowry Payment at the Time of Marriage) | Received Land Inheritance (=1 if yes) | Acreage Logged |
| | (1) | (2) | (3) | (4) | (5) | (6) | (7) | (8) |
| ATT | -0.669*** | 2.476*** | 0.104*** | 0.133*** | 0.029 | -0.374*** | 0.027*** | 0.327*** |
| | (0.176) | (0.189) | (0.009) | (0.013) | (0.023) | (0.104) | (0.009) | (0.067) |
| State FE | Yes | Yes | Yes | Yes | Yes | Yes | Yes | Yes |
| Year FE | Yes | Yes | Yes | Yes | Yes | Yes | Yes | Yes |
| Controls | Yes | Yes | Yes | Yes | Yes | No | No | No |
| Observations | 56,511 | 57,260 | 57,260 | 57,260 | 45,050 | 3,463 | 3,463 | 1,141 |

*Note:* (a) A woman is treated if she is from a state that amended HSA and married after the amendment in her state. (b) Bootstrap (1,000 iterations) standard errors are clustered by state and appear in parentheses. (c) State FE and Year FE (mother's birth year) are added in all columns. (d) Observations used in this table are Hindus (i.e., Hindus, Buddhists, Jains, and Sikhs) only as only they are subject to HSA 1956. (e) columns 1-5 use NFHS data, and columns 6-8 use REDS data. (f) The control variable in columns 1-5 is the survey year, and no controls are included in columns 6-8. (g) The age gap measures the difference between a husband's and a wife's age. (h) Top wealth quintile is an indicator equals one if a woman is married to a top wealth quintile household, and zero otherwise. (i) Husband and wife desire the same fertility outcome equals one if husband and wife have no conflict in their desire for children (i.e., either they both want more children, or they both do not want more children) and equals a zero if husband and wife have different desire for children. (j) Received land inheritance is an indicator that takes one if a woman received a land inheritance after her grandfather's death, and zero otherwise. (k) ***Significant at the 1 percent level, **Significant at the 5 percent level, *Significant at the 10 percent level.

**Table A4**: Heterogenous Effect of HSAA on Bargaining Power by Wealth Quintile

|  | Female Bargaining | | | Bargaining Index |
|---|---|---|---|---|
|  | Say About a Large Purchase (=1 if yes) | Say About Own Health (=1 if yes) | Going to the Market Alone (=1 if yes) |  |
|  | (1) | (2) | (3) | (4) |
| ATT Wealth Quintile 1 (Bottom) | 0.093 | 0.085 | -0.020 | 0.170 |
|  | (0.076) | (0.073) | (0.070) | (0.155) |
| ATT Wealth Quintile 2 | 0.075 | 0.043 | -0.048 | 0.088 |
|  | (0.047) | (0.046) | (0.043) | (0.093) |
| ATT Wealth Quintile 3 | 0.083** | 0.034 | -0.035 | 0.094 |
|  | (0.036) | (0.036) | (0.033) | (0.072) |
| ATT Wealth Quintile 4 | 0.083** | 0.067** | 0.010 | 0.160** |
|  | (0.032) | (0.033) | (0.030) | (0.064) |
| ATT Wealth Quintile 5 (Top) | 0.122*** | 0.115*** | 0.049 | 0.278*** |
|  | (0.035) | (0.035) | (0.036) | (0.072) |
| ATT Wealth Quintile 2- ATT Wealth Quintile 1 | -0.018 | -0.043 | -0.028 | -0.083 |
|  | (0.090) | (0.086) | (0.080) | (0.178) |
| ATT Wealth Quintile 3- ATT Wealth Quintile 1 | -0.011 | -0.051 | -0.015 | -0.076 |
|  | (0.085) | (0.081) | (0.078) | (0.172) |
| ATT Wealth Quintile 4- ATT Wealth Quintile 1 | -0.011 | -0.019 | 0.030 | -0.010 |
|  | (0.082) | (0.080) | (0.076) | (0.168) |
| ATT Wealth Quintile 5- ATT Wealth Quintile 1 | 0.029 | 0.030 | 0.069 | 0.107 |
|  | (0.084) | (0.082) | (0.083) | (0.175) |
| State FE | Yes | Yes | Yes | Yes |
| Year FE | Yes | Yes | Yes | Yes |
| Controls | Yes | Yes | Yes | Yes |
| Observations | 40,152 | 40,153 | 40,408 | 40,152 |

*Note*: (a) A woman is treated if she is from a state that amended HSA and married after the amendment in her state. (b) We use three indicators of bargaining: the woman has a say in large purchases and her own health care and can go to market alone. Bargaining in column 4 is defined as the first principal component of the three decision variables (say about a large purchase, say about own health, and going to market alone). It is normalized to have zero mean and one standard deviation. (c) Bootstrap (1,000 iterations) standard errors are clustered by state and appear in parentheses. (d) State FE and Year FE (mother's birth year) are added in all columns. (e) Control variables are: survey year. (f) Observations used in this table are Hindus (i.e., Hindus, Buddhists, Jains, and Sikhs) only as only they are subject to HSA 1956. (g)***Significant at the 1 percent level, **Significant at the 5 percent level, *Significant at the 10 percent level.

**Table A5:** Test of Common Trend

|  | HFA z-score | | WFA z-score | |
| --- | --- | --- | --- | --- |
|  | (1) | (2) | (3) | (4) |
| Treat × Year | 0.011 | 0.008 | 0.007 | 0.008 |
|  | (0.014) | (0.014) | (0.011) | (0.011) |
| Treat | -0.051 | 0.126 | -0.187 | -0.131 |
|  | (0.179) | (0.171) | (0.135) | (0.137) |
| Year | 0.017*** | 0.056*** | 0.007* | 0.043*** |
|  | (0.006) | (0.009) | (0.007) | (0.008) |
| Survey Year FE | Yes | Yes | Yes | Yes |
| Controls | No | Yes | No | Yes |
| Observations | 10,331 | 10,330 | 10,226 | 10,224 |

*Note*: (a) Treat is a dummy variable that equals one if the state amended the HSA and equals zero otherwise. (b) Year is a linear trend of the mother's year of marriage (c) HFA z-score is the child's height-for-age z-score, and WFA z-score is child's weight-for-age z score. (d) Bootstrap (1,000 iterations) standard errors are clustered by state and appear in parentheses. (e) Survey year FE is added in all columns. (f) Control variables are: mother's age at first birth, mother's age at first birth squared, mother's current age dummies (15-20 years, 21-25 years, 26-30 years, 31-35 years, and 35+ years), child's sex, child age dummies (months), and child's birth order (2nd born, and 3rd+ born). (g) Observations used in this table are Hindus who are subject to HSA 1956. (h) ***Significant at the 1 percent level, **Significant at the 5 percent level, and *Significant at the 10 percent level.

**Table A6:** Possible Selection in HSAA

| | Outcome Variable: HSA Amendment (=1 if yes) | | | |
|---|---|---|---|---|
| | (1) | (2) | (3) | (4) |
| Average Levels in Districts in 1992-93 | | | | |
| HFA | 0.026 | 0.024 | - | - |
| | (0.113) | (0.059) | | |
| WFA | - | - | -0.077 | -0.049 |
| | | | (0.115) | (0.070) |
| Controls | No | Yes | No | Yes |
| Observations | 217 | 217 | 217 | 217 |

*Note:* (a) Sample used in this table are Hindus (who are subject to HSA 1956) from the NFHS-I (1992-93) survey and states that did not amend HSA until 1993. (b) HSA reform is a dummy variable that equals one if a district is subject to the HSA amendment after 1993 (i.e., Karnataka (in 1994) and Maharashtra (in 1994)) and equals zero otherwise. (c) HFA refers to the mean children's height-for-age z-score at the district level. WFA refers to the mean children's weight-for-age z-score at the district level. (d) Bootstrap (1,000 iterations) standard errors are clustered by state and appear in parentheses. (e) Control variables are: mother's age at first birth, mother's age at first birth squared, and mother's current age dummies (15-20 years, 21-25 years, 26-30 years, 31-35 years, and 35+ years). (f) ***Significant at the 1 percent level, **Significant at the 5 percent level, *Significant at the 10 percent level.

**Table A7**: Effect of HSAA on Child Health (Using TWFE)

| | HFA z-score | WFA z-score | Prenatal Inputs | | | | Postnatal Inputs | | Pooled Inputs |
|---|---|---|---|---|---|---|---|---|---|
| | | | Total Prenatal Visits During Pregnancy | Mother Took Iron Supplement During Pregnancy (1= if yes) | Mother Took Tetanus Shot During Pregnancy (=1 if yes) | Delivery at Health Facility (1= if yes) | Postnatal Check Within Two Months of Birth (1= if yes) | Total Vaccination | |
| | (1) | (2) | (3) | (4) | (5) | (6) | (7) | (8) | (9) |
| Treat × After | 0.183*** | 0.276*** | 0.805*** | -0.013 | 0.026* | 0.173*** | 0.139*** | -0.364*** | 0.359*** |
| | (0.065) | (0.049) | (0.119) | (0.016) | (0.014) | (0.018) | (0.026) | (0.110) | (0.048) |
| State FE | Yes | Yes | Yes | Yes | Yes | Yes | Yes | Yes | Yes |
| Year FE | Yes | Yes | Yes | Yes | Yes | Yes | Yes | Yes | Yes |
| State-Year FE | Yes | Yes | Yes | Yes | Yes | Yes | Yes | Yes | Yes |
| Survey Year FE | Yes | Yes | Yes | Yes | Yes | Yes | Yes | Yes | Yes |
| Controls | Yes | Yes | Yes | Yes | Yes | Yes | Yes | Yes | Yes |
| Observations | 56,832 | 56,477 | 50,338 | 33,444 | 33,482 | 56,796 | 23,758 | 56,797 | 23,633 |

*Note:* (a) Treat is a dummy variable that equals one if the mother is from a state that amended HSA before national adoption in 2005 (i.e., reform states) and equals zero otherwise. (b) After is a dummy variable that takes a value of 1 if the mother was married after the reform in her state and takes zero otherwise. (c) HFA z-score is the child's height-for-age z-score, and WFA z-score is child's weight-for-age z-score. (d) Outcome variables in columns 3-6 are prenatal inputs, and Columns 7-8 are postnatal inputs. (e) The mother took the iron supplement, and tetanus shots during pregnancy are available only for the most recent birth (youngest child). (f) Delivery at a health facility refers to childbirth at a health facility. (g) Total vaccination uses children ages 13–59 months, as the recommended age for some is up to one year. (h) Postnatal check within two months of birth is available only for the most recent birth (youngest child) and only in NFHS-2 and NHFS-3. (i) The outcome variable in column 9 is pooled inputs: total prenatal visits during pregnancy, mother took iron supplement during pregnancy, mother took tetanus shot during pregnancy, delivery at a health facility, postnatal check within two months of birth, and child ever vaccinated. We take the first principal component of the standardized inputs and then normalize it to have a mean of zero and a standard deviation of one. (j) Bootstrap (1,000 iterations) standard errors are clustered by state and appear in parentheses. (k) State FE, Year FE (mother's birth year), State-Year FE, and Survey year FE are added in all columns. (l) Control variables are: mother's age at first birth, mother's age at first birth squared, mother's current age dummies (15-20 years, 21-25 years, 26-30 years, 31-35 years, and 35+ years), child's sex, child age dummies (months), and indicators of child's birth order (2nd born and 3rd+ born). (m) Observations used in this table are Hindus (i.e., Hindus, Buddhists, Jains, and Sikhs) only as they are subject to HSA 1956. (n) ***Significant at the 1 percent level, **Significant at the 5 percent level, and *Significant at the 10 percent level.

Table A8: Heterogeneous Effect of HSAA on Child Health by Birth Order

| | HFA z-score | WFA z-score | Prenatal Inputs | | | Postnatal Inputs | | | Pooled Inputs |
|---|---|---|---|---|---|---|---|---|---|
| | | | Total Prenatal Visits During Pregnancy | Mother Took Iron Supplement During Pregnancy (1= if yes) | Mother Took Tetanus Shot During Pregnancy (=1 if yes) | Delivery at Health Facility (1= if yes) | Postnatal Check Within Two Months of Birth (1= if yes) | Total Vaccination | |
| | (1) | (2) | (3) | (4) | (5) | (6) | (7) | (8) | (9) |
| Treat × After × 2nd Child | -0.021 | 0.038 | -0.231** | -0.012 | 0.007 | 0.030** | -0.040* | 0.089 | -0.096** |
| | (0.047) | (0.035) | (0.094) | (0.012) | (0.009) | (0.012) | (0.021) | (0.067) | (0.038) |
| Treat × After × 3rd+ Child | 0.003 | 0.046 | -0.293** | 0.027 | 0.035*** | 0.050*** | -0.074*** | 0.446*** | -0.186*** |
| | (0.067) | (0.050) | (0.132) | (0.018) | (0.013) | (0.018) | (0.028) | (0.102) | (0.055) |
| Treat × After | 0.191*** | 0.252*** | 0.955*** | -0.015 | 0.015 | 0.151*** | 0.164*** | -0.492*** | 0.420*** |
| | (0.071) | (0.054) | (0.132) | (0.018) | (0.015) | (0.020) | (0.027) | (0.120) | (0.051) |
| 2nd Child (=1 if yes) | -0.246*** | -0.227*** | -0.898*** | -0.066*** | -0.047*** | -0.186*** | -0.080*** | -0.600*** | -0.300*** |
| | (0.021) | (0.017) | (0.041) | (0.008) | (0.006) | (0.005) | (0.008) | (0.039) | (0.018) |
| 3rd+ Child (=1 if yes) | -0.636*** | -0.549*** | -2.317*** | -0.195*** | -0.158*** | -0.398*** | -0.157*** | -1.593*** | -0.676*** |
| | (0.026) | (0.020) | (0.052) | (0.009) | (0.007) | (0.007) | (0.010) | (0.050) | (0.022) |
| State FE | Yes | Yes | Yes | Yes | Yes | Yes | Yes | Yes | Yes |
| Year FE | Yes | Yes | Yes | Yes | Yes | Yes | Yes | Yes | Yes |
| State-Year FE | Yes | Yes | Yes | Yes | Yes | Yes | Yes | Yes | Yes |
| Survey Year FE | Yes | Yes | Yes | Yes | Yes | Yes | Yes | Yes | Yes |
| Controls | Yes | Yes | Yes | Yes | Yes | Yes | Yes | Yes | Yes |
| Observations | 56,832 | 56,477 | 50,338 | 33,444 | 33,482 | 56,796 | 23,758 | 41,121 | 23,633 |

*Note:* (a) Treat is a dummy variable that equals one if the mother is from a state that amended HSA before national adoption in 2005 (i.e., reform states) and equals zero otherwise. (b) After is a dummy variable that takes a value of 1 if the mother was married after the reform in her state and takes zero otherwise. (c) HFA z-score is the child's height-for-age z-score, and WFA z-score is child's weight-for-age z-score. (d) The outcome variables in columns 5-8 are prenatal, and 7-8 are postnatal inputs. (e) The mother took the iron supplement, and tetanus shots during pregnancy are available only for the most recent birth (youngest child). (f) Delivery at a health facility refers to childbirth at a health facility. (g) Total vaccination uses children ages 13–59 months, as the recommended age for some is up to one year. (h) Postnatal check within two months of birth is available only for the most recent birth (youngest child) and only in NFHS-2 and NHFS-3. (i) The outcome variable in column 9 is pooled inputs: total prenatal visits during pregnancy, mother took iron supplement during pregnancy, mother took tetanus shot during pregnancy, delivery at a health facility, postnatal check within two months of birth, and child ever vaccinated. We take the first principal component of the standardized inputs and then normalize it to have a mean of zero and a standard deviation of one. (j) Bootstrap (1,000 iterations) standard errors are clustered by state and appear in parentheses. (k) State FE, Year FE (mother's birth year), State-Year FE, and Survey year FE are added in all columns. (l) Control variables are: mother's age at first birth, mother's age at first birth squared, mother's current age dummies (15-20 years, 21-25 years, 26-30 years, 31-35 years, and 35+ years), child's sex, and child age dummies (months). (m) 2nd child is an indicator for children whose birth order is 2, and 3rd+ child is an indicator for children whose birth order is 3 or higher. (n) Observations used in this table are Hindus (i.e., Hindus, Buddhists, Jains, and Sikhs) only as they are subject to HSA 1956. (o) ***Significant at the 1 percent level, **Significant at the 5 percent level, and *Significant at the 10 percent level.

Table A9: Heterogeneous Treatment Effect of HSAA on Child Health by Child's Sex

|  | HFA z-score | WFA z-score | Prenatal Inputs | | | | Postnatal Inputs | | Total Vaccination | Pooled Inputs |
|---|---|---|---|---|---|---|---|---|---|---|
|  |  |  | Total Prenatal Visits During Pregnancy | Mother Took Iron Supplement During Pregnancy (1= if yes) | Mother Took Tetanus Shot During Pregnancy (=1 if yes) | Delivery at Health Facility (1= if yes) | Postnatal Check Within Two Months of Birth (1= if yes) |  |  |  |
|  | (1) | (2) | (3) | (4) | (5) | (6) | (7) | (8) | (9) |
| Treat × After × Girl | 0.021 | 0.064** | 0.021 | 0.022** | 0.022*** | 0.001 | -0.005 | 0.319*** | 0.042 |
|  | (0.041) | (0.032) | (0.085) | (0.011) | (0.008) | (0.010) | (0.016) | (0.059) | (0.033) |
| Treat × After | 0.149** | 0.187*** | 0.730*** | -0.027* | 0.016 | 0.149*** | 0.111*** | -0.513*** | 0.279*** |
|  | (0.062) | (0.048) | (0.117) | (0.015) | (0.012) | (0.018) | (0.024) | (0.111) | (0.045) |
| Girl (=1 if yes) | 0.041*** | -0.006 | -0.088*** | -0.009* | -0.018*** | -0.015*** | -0.025*** | -0.304*** | -0.058*** |
|  | (0.015) | (0.012) | (0.025) | (0.005) | (0.005) | (0.004) | (0.005) | (0.031) | (0.011) |
| State FE | Yes | Yes | Yes | Yes | Yes | Yes | Yes | Yes | Yes |
| Year FE | Yes | Yes | Yes | Yes | Yes | Yes | Yes | Yes | Yes |
| State-Year FE | Yes | Yes | Yes | Yes | Yes | Yes | Yes | Yes | Yes |
| Survey Year FE | Yes | Yes | Yes | Yes | Yes | Yes | Yes | Yes | Yes |
| Controls | Yes | Yes | Yes | Yes | Yes | Yes | Yes | Yes | Yes |
| Observations | 57,224 | 56,867 | 50,702 | 33,807 | 33,845 | 57,188 | 24,054 | 41,121 | 23,928 |

*Note:* (a) Treat is a dummy variable that equals one if the mother is from a state that amended HSA before national adoption in 2005 (i.e., reform states) and equals zero otherwise. (b) After is a dummy variable that takes a value of 1 if the mother was married after the reform in her state and takes zero otherwise. (c) HFA z-score is the child's height-for-age z-score, and WFA z-score is child's weight-for-age z-score. (d) Outcome variables in columns 3-6 are prenatal inputs, and columns 7-8 are postnatal inputs. (e) The mother took the iron supplement, and tetanus shots during pregnancy are available only for the most recent birth (youngest child). (f) Delivery at a health facility refers to childbirth at a health facility. (g) Total vaccination uses children ages 13–59 months, as the recommended age for some is up to one year. (h) Postnatal check within two months of birth is available only for the most recent birth (youngest child) and only in NFHS-2 and NHFS-3. (h) Postnatal check within two months of birth is available only for the most recent birth (youngest child) and only in NFHS-2 and NHFS-3. (i) The outcome variable in column 9 is pooled inputs: total prenatal visits during pregnancy, mother took iron supplement during pregnancy, mother took tetanus shot during pregnancy, delivery at a health facility, postnatal check within two months of birth, and child ever vaccinated. We take the first principal component of the standardized inputs and then normalize it to have a mean of zero and a standard deviation of one. (j) Bootstrap (1,000 iterations) standard errors are clustered by state and appear in parentheses. (k) State FE, Year FE (mother's birth year), State-Year FE, and Survey year FE are added in all columns. (l) Control variables are: mother's age at first birth, mother's age at first birth squared, mother's current age dummies (15-20 years, 21-25 years, 26-30 years, 31-35 years, and 35+ years), child age dummies (months), and indicators of child's birth order (2nd born and 3rd+ born). (m) Observations used in this table are Hindus (i.e., Hindus, Buddhists, Jains, and Sikhs) only as they are subject to HSA 1956. (n) ***Significant at the 1 percent level, **Significant at the 5 percent level, and *Significant at the 10 percent level.

**Table A10**: Heterogeneous Effect of HSAA on Child Health by Birth Order and Sex of the Children

|  | HFA z-score | WFA z-score | Prenatal Inputs ||||| Postnatal Inputs || Total Vaccination | Pooled Inputs |
|---|---|---|---|---|---|---|---|---|---|
|  |  |  | Total Prenatal Visits During Pregnancy | Mother Took Iron Supplement in Pregnancy (1= if yes) | Mother Took Tetanus Shot During Pregnancy (=1 if yes) | Delivery at Health Facility (1= if yes) | Postnatal Check Within Two Months of Birth (1= if yes) |  |  |
|  | (1) | (2) | (3) | (4) | (5) | (6) | (7) | (8) | (9) |
| Treat × After × 2nd Child × Girl | 0.083 | -0.016 | 0.001 | 0.047* | 0.019 | 0.028 | 0.052 | 0.321** | 0.086 |
|  | (0.095) | (0.076) | (0.193) | (0.025) | (0.018) | (0.023) | (0.039) | (0.132) | (0.073) |
| Treat × After × 3rd+ Child × Girl | 0.191 | 0.113 | 0.063 | 0.048 | 0.019 | 0.016 | 0.016 | 0.268 | 0.040 |
|  | (0.120) | (0.094) | (0.221) | (0.032) | (0.024) | (0.032) | (0.050) | (0.187) | (0.092) |
| Treat × After × 2nd Child | -0.062 | 0.047 | -0.232* | -0.034** | -0.001 | 0.016 | -0.065** | -0.053 | -0.137*** |
|  | (0.066) | (0.050) | (0.134) | (0.017) | (0.012) | (0.016) | (0.027) | (0.092) | (0.053) |
| Treat × After × 3rd+ Child | -0.087 | -0.005 | -0.324* | 0.005 | 0.026 | 0.043* | -0.082** | 0.331** | -0.205*** |
|  | (0.088) | (0.068) | (0.165) | (0.023) | (0.017) | (0.024) | (0.038) | (0.129) | (0.072) |
| Treat × After × Girl | -0.060 | 0.031 | -0.029 | -0.009 | 0.002 | -0.014 | -0.025 | 0.097 | -0.015 |
|  | (0.063) | (0.050) | (0.134) | (0.017) | (0.012) | (0.015) | (0.025) | (0.085) | (0.049) |
| 2nd Child × Girl | -0.118*** | -0.058* | -0.074 | -0.008 | -0.009 | -0.020* | -0.001 | -0.218*** | -0.032 |
|  | (0.041) | (0.032) | (0.077) | (0.015) | (0.012) | (0.011) | (0.015) | (0.077) | (0.033) |
| 3rd+ Child × Girl | -0.112*** | -0.094*** | -0.153** | -0.014 | -0.033*** | -0.015* | 0.006 | -0.309*** | -0.061** |
|  | (0.038) | (0.029) | (0.062) | (0.013) | (0.011) | (0.009) | (0.013) | (0.072) | (0.028) |
| Treat × After | 0.220*** | 0.237*** | 0.969*** | -0.011 | 0.014 | 0.158*** | 0.176*** | -0.540*** | 0.426*** |
|  | (0.078) | (0.059) | (0.148) | (0.019) | (0.016) | (0.021) | (0.030) | (0.126) | (0.058) |
| 2nd Child (=1 if yes) | -0.189*** | -0.199*** | -0.862*** | -0.062*** | -0.043*** | -0.176*** | -0.080*** | -0.495*** | -0.284*** |
|  | (0.030) | (0.023) | (0.055) | (0.010) | (0.008) | (0.008) | (0.011) | (0.054) | (0.024) |
| 3rd+ Child (=1 if yes) | -0.581*** | -0.504*** | -2.244*** | -0.188*** | -0.143*** | -0.391*** | -0.160*** | -1.1446*** | -0.646*** |
|  | (0.031) | (0.025) | (0.059) | (0.011) | (0.009) | (0.009) | (0.012) | (0.060) | (0.026) |
| Girl (=1 if yes) | 0.122*** | 0.051** | -0.001 | 0.001 | 0.001 | -0.003 | -0.026** | -0.103* | -0.019 |
|  | (0.029) | (0.023) | (0.055) | (0.011) | (0.008) | (0.008) | (0.011) | (0.054) | (0.024) |
| State FE | Yes | Yes | Yes | Yes | Yes | Yes | Yes | Yes | Yes |
| Year FE | Yes | Yes | Yes | Yes | Yes | Yes | Yes | Yes | Yes |
| State-Year FE | Yes | Yes | Yes | Yes | Yes | Yes | Yes | Yes | Yes |
| Survey Year FE | Yes | Yes | Yes | Yes | Yes | Yes | Yes | Yes | Yes |
| Controls | Yes | Yes | Yes | Yes | Yes | Yes | Yes | Yes | Yes |
| Observations | 56,832 | 56,477 | 50,338 | 33,444 | 33,482 | 56,796 | 23,758 | 41,121 | 23,633 |

*Note:* (a) Treat is a dummy variable that equals one if the mother is from a state that amended HSA before national adoption in 2005 (i.e., reform states) and equals zero otherwise. (b) After is a dummy variable that takes a value of 1 if the mother was married after the reform in her state and takes zero otherwise. (c) HFA z-score is the child's height-for-age z-score, and WFA z-score is child's weight-for-age z-score. (d) Outcome variables in columns 3-6 are prenatal inputs, and Columns 7-8 are postnatal inputs. (e) The mother took the iron supplement, and tetanus shots during pregnancy are available only for the most recent birth (youngest child). (f) Delivery at a health facility refers to childbirth at a health facility. (g) Total vaccination uses children ages 13–59 months, as the recommended age for some is up to one year. (h) Postnatal check within two months of birth is available only for the most recent birth (youngest child) and only in NFHS-2 and NFHS-3. (i) The outcome variable in column 9 is pooled inputs: total prenatal visits during pregnancy, mother took iron supplement during pregnancy, mother took tetanus shot during pregnancy, delivery at a health facility, postnatal check within two months of birth, and child ever vaccinated. We take the first principal component of the standardized inputs and then normalize it to have a mean of zero and a standard deviation of one. (j) Bootstrap (1,000 iterations) standard errors are clustered by state and appear in parentheses. (k) State FE, Year FE (mother's birth year), State-Year FE, and Survey year FE are added in all columns. (l) Control variables are: mother's age at first birth, mother's age at first birth squared, mother's current age dummies (15-20 years, 21-25 years, 26-30 years, 31-35 years, and 35+ years), child's sex, child age dummies (months), and child's birth order (2nd born, and 3rd+ born). (m) 2nd child is an indicator for children whose birth order is 2, and 3rd+ child is an indicator for children whose birth order is 3 or higher. (n) Girl is an indicator for the children who are girls. (O) Observations used in this table are Hindus (i.e., Hindus, Buddhists, Jains, and Sikhs) only as they are subject to HSA 1956. (P) ***Significant at the 1 percent level, **Significant at the 5 percent level, and *Significant at the 10 percent level.

**Table A11**: Effect of HSAA on Child Health with Alternative Samples

|  | Adding Union Territories and the Northeastern States | | Adding Union Territories, Northeastern States, West Bengal, and Jammu and Kashmir | | Adding Union Territories, Northeastern States, West Bengal, Jammu and Kashmir, and Kerala | |
|---|---|---|---|---|---|---|
|  | HFA z-score | WFA z-score | HFA z-score | WFA z-score | HFA z-score | WFA z-score |
|  | (1) | (2) | (3) | (4) | (5) | (6) |
| ATT | 0.197*** | 0.284*** | 0.193*** | 0.283*** | 0.212* | 0.266*** |
|  | (0.068) | (0.051) | (0.067) | (0.050) | (0.110) | (0.065) |
| Observations | 58,277 | 57,904 | 61,505 | 61,116 | 62,705 | 62,311 |
| State FE | Yes | Yes | Yes | Yes | Yes | Yes |
| Year FE | Yes | Yes | Yes | Yes | Yes | Yes |
| Controls | Yes | Yes | Yes | Yes | Yes | Yes |

Note: (a) This table reports the average treatment effect on the treated (ATT) of HSAA on child health outcomes. (b) A woman is treated if she is from a state that amended HSA and married after the amendment in her state. (c) The HFA z-score is the child's height-for-age z-score, and the WFA z-score is the child's weight-for-age z score. (d) Bootstrap (1,000 iterations) standard errors are clustered by state and appear in parentheses. (e) State FE and Year FE (women's birth year) are added in all columns. (f) Control variables are the child's sex, age (in months), birth order, household size (number of household members), and survey year. (g) Observations used in this table are Hindus (i.e., Hindus, Buddhists, Jains, and Sikhs) only as they are subject to HSA 1956. (h) ***Significant at the 1 percent level, **Significant at the 5 percent level, and *Significant at the 10 percent level.

# Online Appendix B

Panel A. HFA Trends in Reform and Non-reform States Before Reforms Enacted

Panel B. WFA Trends in Reform and Non-reform States Before Reforms Enacted

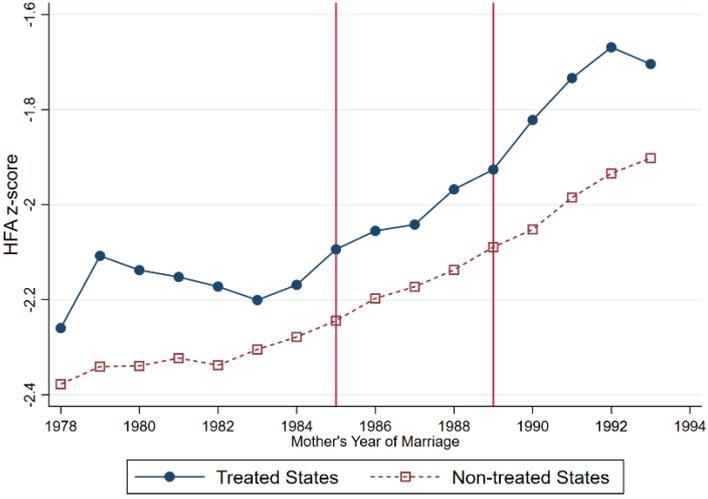
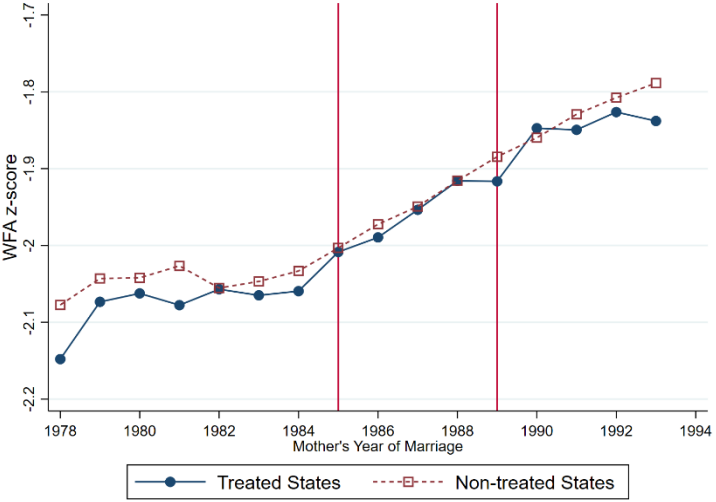

**Figure B1:** HFA and WFA Trends in Reform and the Non-reform States Before Enacted Reforms